\RequirePackage{amsmath}
\documentclass[epj]{svjour}
\usepackage{amssymb}
\usepackage{textcomp}
\usepackage{graphicx}
\usepackage{grffile}
\usepackage{color}
\usepackage[caption=false]{subfig}
\usepackage{bm}
\usepackage{tabularx}
\definecolor{Green}{rgb}{0.000000,0.392157,0.000000}
\definecolor{Purple}{rgb}{0.5,0.1,0.9}
\definecolor{Blue}{rgb}{0.0,0.0,1.0}
\usepackage{placeins}
\newcommand{\stl}[1]{ \mbox{ $\!\stackrel{ \rule[-1.5pt]{0.5pt}{0.65ex} \overline{\hphantom{\mbox{$\displaystyle #1$}}} \rule[-1.5pt]{0.5pt}{0.65ex} }{#1}\!$ } }
\newcommand{\Tr}{\text{Tr}}
\usepackage{mathrsfs}
\usepackage{soul}

\usepackage[USenglish]{babel}
\begin{document}
\title{Shear banding in nematogenic fluids with oscillating orientational dynamics}
\titlerunning{Shear banding in nematogenic fluids}
\author{R.~Lugo-Frias, H.~Reinken \and S.~H.~L.~Klapp
}                     
\institute{Institut f\"ur Theoretische Physik, Technische Universit\"at Berlin, Hardenbergstrasse 36, 10623 Berlin, Germany}
\date{Received: date / Revised version: date}
%
\abstract{
We investigate the occurrence of shear banding in nematogenic fluids under planar Couette flow, based on mesoscopic dynamical equations for the orientational order parameter and the shear stress. We focus on parameter values where the sheared homogeneous system exhibits regular oscillatory orientational dynamics, whereas the equilibrium system is either isotropic (albeit close to the isotropic--nematic transition) or deep in its nematic phase. The numerical calculations are restricted to spatial variations in shear gradient direction. We find several new types of shear banded states characterized by regions with regular oscillatory orientational dynamics. In all cases shear banding is accompanied by a non--monotonicity of the flow curve of the homogeneous system; however, only in the case of the initially isotropic system this curve has the typical $S$--like shape. We also analyze the influence of different orientational boundary conditions and of the spatial correlation length.
} 
\maketitle
\section{Introduction}
\label{Sec:01}
The emergence of banded structures in complex fluids under shear flow is a paradigmatic example of an instability in a correlated soft matter system far from equilibrium. Above
a critical value of the applied shear rate (or shear stress), the formerly homogeneous system becomes unstable and separates into macroscopic bands with different {\em local} shear rates (stresses), see Refs.~\cite{Fielding2007,Dhont2008} for recent reviews. Typical systems where shear band formation has been observed experimentally are wormlike micelles~\cite{LopezGonzalez2004}, liquid--crystalline polymers~\cite{Decruppe1995}, colloidal suspensions~\cite{Chen1992}, but also non--ergodic soft systems such as glasses~\cite{Chikkadi2014}.
In all cases, the flow leads to reorganization of the fluid's microstructure which then feeds back into the flow field. This eventually leads to a non--monotonicity of the flow curve, that is, 
the relation between shear stress and shear rate. In that sense, non--monotonic flow curves are signatures of shear banding. Theoretically, shear banding (and the related vorticity banding~\cite{Goveas2001}) has been studied mainly via continuum 
models. An important example is the diffusive (non--local) Johnson-Segelman (DJS) model~\cite{Spenley1996,Olmsted2000} for 
shear thinning systems, {\em i.e.}, systems in which the viscosity decreases with the stress, which form shear bands along the gradient direction.
Moreover, particle resolved simulations~\cite{Tao2005,Ripoll2008} have revealed insight into microscopic mechanisms accompanying shear thinning and shear thickening systems.

In the present paper we focus on shear banding in nematogenic fluids whose anisotropic constituents can arrange into orientationally ordered, yet transitionally disordered states. Prominent examples are wormlike micelles and colloidal suspensions of rod--like particles, both of which display flow--induced spatial instabilities in experiments. From the theoretical side,
the shear--induced behavior of nematogenic fluids has been intensely studied on the basis of nonlinear equations for the dynamics of the orientational order parameter the so--called $\mathbf{Q}$--tensor, with coupling to the concentration~\cite{Olmsted1999} or for dense systems deep in the nematic phase~\cite{Hess1994,Rienacker2000,Rienacker2002,Strehober2013}.
Contrary to the DJS model, the $\mathbf{Q}$--tensor models allow to investigate directly the impact of shear on the structure (on a coarse--grained, order parameter level), from which the shear stress can then be derived by additional relations. 

Already for homogeneous systems, these $\mathbf{Q}$--tensor models predict 
many interesting effects such as the shear--induced shift of the isotropic--nematic transition~\cite{Olmsted1991}, and the occurrence of dynamical states with regular or even chaotic oscillatory motion of the
nematic director~\cite{Hess1994,Rienacker2002}. Indications of such {\em time--dependent} dynamical states under steady shear flow have also been observed in many--particle simulations~\cite{Tao2005,Ripoll2008}
 and in experiments~\cite{Roux1995,Lettinga2005}.
In addition, $\mathbf{Q}$--tensor models have been used to explore spatial inhomogeneities, yielding shear banding between differently steady (aligned) states
close to the isotropic--nematic transition~\cite{Olmsted1999}
and in parameter regimes where the dynamics of the homogeneous system is chaotic~\cite{Chakraborty2010}.

In the present article we consider (as in~\cite{Chakraborty2010}) systems at constant concentration where deviations from the applied 
flow profile are taken into account in the Stokesian limit. Our purpose
is to extend the earlier studies~\cite{Chakrabarti2004,Das2005} on shear banding in nematogenic fluids along the gradient direction in several ways. 
First, we focus on parameters where the homogeneous systems exhibits {\em regular} oscillatory dynamics. These are, on the one hand, initially ({\em i.e.}, at $\dot\gamma=0$)
isotropic systems at low tumbling parameters
and high shear rates, which display wagging motion~\cite{Strehober2013} and, on the other hand, initially nematic systems where dynamical modes such as tumbling and kayaking are well established (see e.g. ref.~\cite{Rienacker2002}).
Second, we consider both, homogeneous and inhomogeneous systems (with inhomogeneities in gradient direction) in order to identify the signatures of shear banding 
in the homogeneous flow curves. In fact, only one of the considered systems is characterized by a "van--der--Waals" like flow curve (which is familiar, e.g., within the DJS model~\cite{Olmsted2000}); 
the other ones rather exhibit discontinuities.
Third, we explore systematically the impact of different orientational boundary conditions and different  correlation lengths. We show, in particular, that appropriate boundaries can {\em induce} banding
in otherwise homogeneous states. Further, we discuss the consequences for the stress of the system in the banded state.

The article is organized as follows. In sect.~\ref{Sec:02} we introduce the set of dynamical equations for spatially inhomogeneous, anisotropic fluids at constant density (with inhomogeneities along
the direction of the shear gradient). Numerical results are presented in sect.~\ref{Sec:03}. There, we first discuss (sect.~\ref{SubSec:03.1}) spatially homogeneous systems sheared from isotropic or nematic states and obtain the corresponding flow curves. Section~\ref{SubSec:03.2} is then devoted 
to the appearance of shear bands and the impact of boundary conditions. This work finishes with concluding remarks and an outlook in sect.~\ref{Sec:04}.

\section{Theoretical framework}
\label{Sec:02}
In the present work we consider systems of uniaxial, rigid rod-like particles (such as suspensions of {\em fd}--viruses~\cite{Dhont2002}) whose orientation is characterized by the unit vector ${\mathbf{u}}_i$ parallel to the
symmetry axes of particle $i$.
Following earlier studies within the Doi-Hess theory, the dynamics of the many-particle system is described by the space-- and time--dependent tensorial order parameter $\mathbf{Q}(\mathbf{r},t)$ (thus, density variations are neglected)~\cite{deGennes1993}. This second-rank $\mathbf{Q}$--tensor is defined as
\begin{equation}
\label{Eq:Q-tensor}
{\mathbf{Q}}(\mathbf{r},t) = \sqrt{\frac{15}{2}} \int\limits_{\mathbb{S}^2} \stl{{\mathbf{u}}{\mathbf{u}}} f(\mathbf{r},{\mathbf{u}},t)\, d {\mathbf{u}}\,,
\end{equation}
where $f(\mathbf{r},{\mathbf{u}},t)$ is the space-- and time--dependent orientational distribution function (for a microscopic definition, see, e.g.,~\cite{Strehober2013}), and $\stl{\mathbf{x}}$ stands for the symmetric traceless part of the tensor $\mathbf{x}$. Explicitly, one has
$\stl{x}_{\mu \nu} = ({x}_{\mu \nu} + {x}_{\nu \mu})/2 - \Tr(x)\mathbb{I}_{\mu \nu}/3$, where $\mu$ and $\nu$ are Cartesian indices, $\mathbb{I}_{\mu \nu}$ is the unit matrix,
and $\Tr$ denotes the trace. Thus, the $\mathbf{Q}$--tensor is, by definition, a symmetric traceless tensor. For the special case of uniaxial nematic phases, the $\mathbf{Q}$--tensor reduces to the form
${\mathbf{Q}} = \mu_3 3/2 \langle\stl{\mathbf{nn}}\rangle$, where $\mathbf{n}$ is the system--averaged nematic director, {\em i.e.}, the eigenvector related to the largest eigenvalue $\mu_3$. 
In equilibrium, $\mu_3$ is proportional to the well--known Maier--Saupe order parameter, $\mu_3 = \sqrt{10/3} S$ where $S\equiv\langle P_2({\mathbf{u}} \cdot \mathbf{n})\rangle$ and $P_2$ denotes the second Legendre polynomial~\cite{Hess2015}. 
\subsection{Mesoscopic dynamical equations}
\label{SubSec:02.1}
In equilibrium, the orientational order of systems of rod-like particles is controlled by the temperature $T$ (typical for molecular fluids) 
and/or by the number density (concentration) $\rho$; the latter case is characteristic of colloidal suspensions of {\em fd}--viruses~\cite{Dhont2002}. On a mesoscopic ($\mathbf{Q}$--tensor) level, the stability of homogeneous (isotropic or nematic) phases is governed by the Landau--type orientational free energy density
\begin{align}
\label{Eq:Landau-Energy}
\mathscr{F}^{or}_{h} = A\!\left({\mathbf{Q}}\!:\!{\mathbf{Q}} \right) - B\! \left( {\mathbf{Q}} \!\cdot\!{\mathbf{Q}} \right)\!:\!{\mathbf{Q}} + C\! \left({\mathbf{Q}} \!:\!{\mathbf{Q}}\right)^2 \,,
\end{align}
where $A, B$ and $C$ are dimensionless coefficients. These can be related to system parameters such as 
$\rho$ and the molecular aspect ratio $\kappa$ as shown, e.g., in~\cite{LugoFrias2016}.

For spatially inhomogeneous phases, an additional contribution to the free energy arises due to the energy cost associated with local deformations of the alignment field. For uniaxial nematic order, an expansion 
of this {\em elastic} energy in terms of the director $\mathbf{n}$ was derived by Oseen~\cite{Oseen1933} and Frank~\cite{Frank1958}. 
Rewriting the expression in terms of the full $\mathbf{Q}$--tensor yields
\begin{align}
\label{Eq:Elastic-Energy}
\mathscr{F}^{or}_{ih} = \frac{1}{2} \xi^2 \, \Tr((\nabla \mathbf{Q})\cdot(\nabla \mathbf{Q})) \,,
\end{align}
where $\xi$ is the elastic correlation length, which is related to the pitch length of the Frank elastic theory~\cite{deGennes1993,Hess2015}. On a microscopical level, $\xi$ is related to the direct correlation function of the fluid~\cite{Singh1985,Singh1986,Singh1987}.

The presence of shear strongly affects the overall orientational ordering due to the competition between flow-induced effects on {\em individual} molecular orientations (such as Jeffery orbits~\cite{Taylor1923})
and the relaxation of the entire system towards equilibrium (governed by the free energy density). Moreover, in {\em inhomogeneous} systems 
(with space-dependent $\mathbf{Q}$--tensor) the orientational dynamics feedback into 
the flow profile through the system's stress tensor. Within the Doi-Hess theory this interplay between flow (characterized through the velocity profile $\mathbf{v}(\mathbf{r})$) and
orientational ordering is described by a set of coupled non--linear equations for $\mathbf{Q}$ and $\mathbf{v}$ which can be derived on the basis of non-equilibrium irreversible thermodynamics~\cite{Hess1975,Pardowitz1980}. Disregarding non--convective flow (corresponding to fourth--order derivatives of the alignment), these equations are given by
\begin{align}
\label{Eq:Hess-Qtensor}
\frac{d \mathbf{Q}}{d t} &= \mathbf{H}(\mathbf{Q},\mathbf{v})+\frac{\xi^2}{\tau_q} \nabla^2 \mathbf{Q} \,,\\
\label{Eq:Hess-Stensor}
\rho \frac{d \mathbf{v}}{d t} &= \nabla \mathbf{T}\,.
\end{align}
Mathematically, eq.~(\ref{Eq:Hess-Qtensor}) is a parabolic equation with $\mathbf{H}(\mathbf{Q},\mathbf{v})$ acting as a source term. 
From a physical point of view, it describes the dynamical evolution of the order parameter including a diffusive term $\propto \nabla^2 \mathbf{Q}$ due to the elastic energy in eq.~(\ref{Eq:Elastic-Energy}). The source term is
given by~\cite{Heidenreich2009,Borgmeyer1995}
\begin{align}
\label{Eq:Hess-Source}
\mathbf{H}(\mathbf{Q},\mathbf{v}) =2\!\stl{\mathbf{\Omega}\!\cdot\!\mathbf{Q}}\! + 2 \sigma \!\stl{\mathbf{\Gamma}\!\cdot\!\mathbf{Q}}\! - \sqrt{2}\frac{\tau_{qp}}{\tau_q} \mathbf{\Gamma} - \frac{1}{\tau_q}\mathbf{\Phi}'\, ,
\end{align}
which describes the interplay between flow, entering via the vorticity $\mathbf{\Omega} = 1/2(\nabla\mathbf{v}^T - \nabla\mathbf{v})$ and the deformation rate
$\mathbf{\Gamma} = 1/2(\nabla\mathbf{v}^T + \nabla\mathbf{v})$, and relaxation towards equilibrium entering via the free energy derivative
\begin{align}
\label{Eq:Hess-DerEnergy}
\mathbf{\Phi}' = \frac{\partial \mathscr{F}^{or}_{h}}{\partial \mathbf{Q}}\,.
\end{align}

Equation~(\ref{Eq:Hess-Source}) further involves the relaxational times ${\tau_{qp}}$ and ${\tau_q}$~\cite{Hess1975}. 
As seen from eqs.~(\ref{Eq:Hess-Qtensor}) and~(\ref{Eq:Hess-Source}), the ratio of these two times [appearing as a prefactor of $\mathbf{\Gamma}$ in eq.~(\ref{Eq:Hess-Source})]
quantifies the perturbation of the system in the absence of orientational order. It is thus convenient to introduce, as a  coupling parameter, the so--called
tumbling parameter $\lambda = - {\tau_{qp}}/{\tau_q}$ which is related to the molecular aspect (length--to--breadth) ratio $\kappa$ viz (see ref.~\cite{Hess1976})
\begin{align}
\label{Eq:Tumbling}
\lambda = -\frac{\tau_{qp}}{\tau_q} = \sqrt{\frac{3}{5}} \frac{\kappa^2 - 1}{\kappa^2 + 1}\,.
\end{align}
Since $\lambda$ depends only on the aspect ratio, it follows from eq.~(\ref{Eq:Tumbling}) that the MDHT is suitable to study spherical particles ($\lambda = 0$), disk--like particles ($\lambda < 0$) and rod--like particles ($\lambda > 0$). As stated before, our focus is on rod--like particles, thus, we consider positive values of the tumbling parameter.

\begin{figure}
\centering
\includegraphics[height=4.5cm,keepaspectratio]{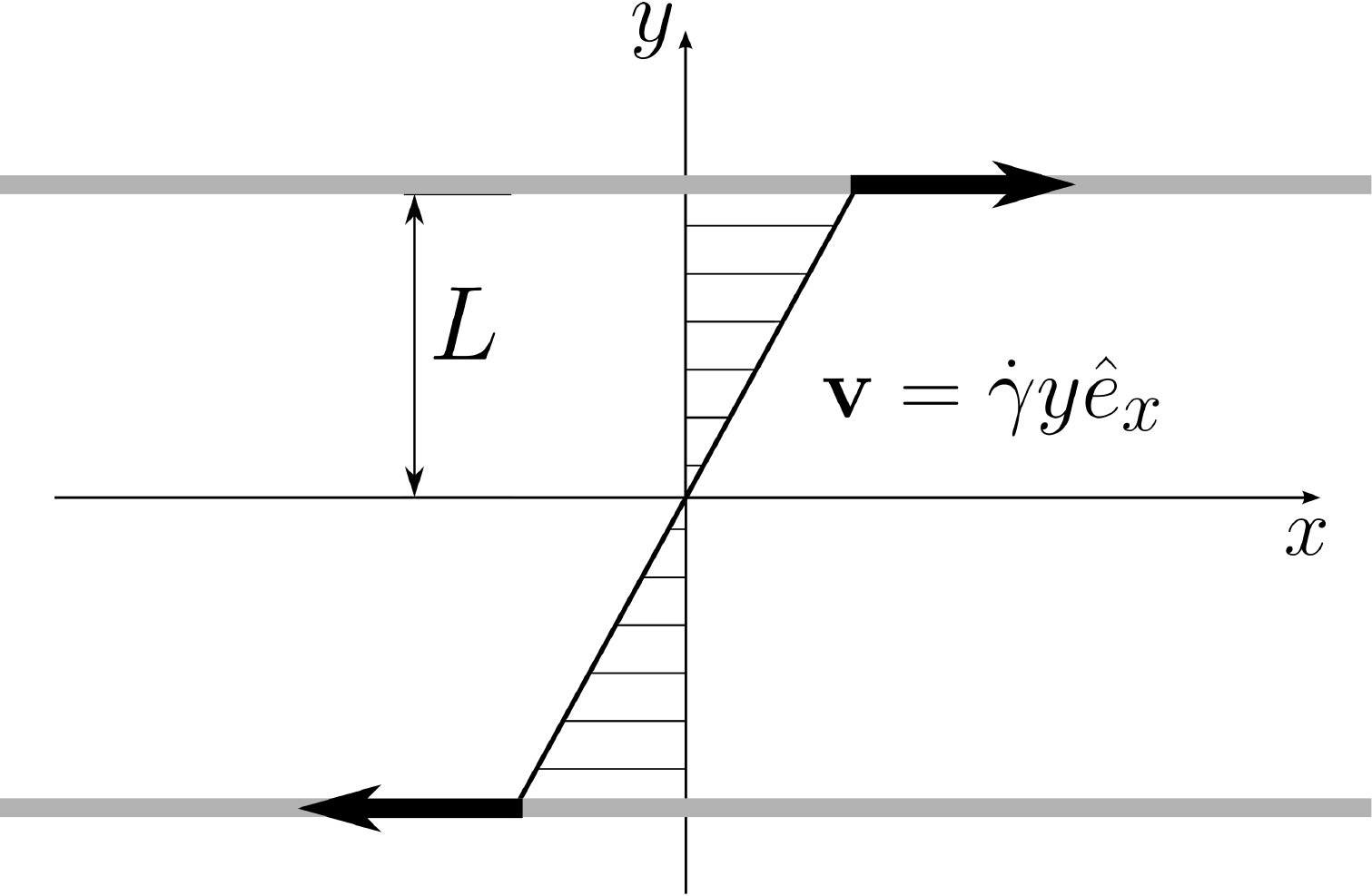}
\caption{Sketch of a planar Couette flow. The rod--like suspension is enclosed between two infinite parallel plates at $y = \pm L$ moving along the $x$--axis with velocities $v_x= \pm L\dot{\gamma}$.}
\label{Fig:Sketch}
\end{figure}

The second mesoscopic equation~(\ref{Eq:Hess-Stensor}) is the usual momentum balance equation~\cite{deGroot1983}; it describes how the flow field changes due to spatial variations of the stress tensor $\mathbf{T}$. 
We here consider a planar Couette flow (see fig.~\ref{Fig:Sketch}) where the fluid is confined between two infinitely extended, parallel plates (separated by a distance $2L$ along the $y$--direction) moving in opposite directions. The flow profile (for a Newtonian fluid) is then given by $\mathbf{v}(\mathbf{r}) = \dot{\gamma}y\hat{\mathbf{e}}_x$, with $\dot{\gamma}$ the shear rate.

The full stress tensor can be written as $\mathbf{T}^{tot}=-p\,\mathbb{I}+\mathbf{T}^{asy}+\stl{\mathbf{T}}\!$, where the first term represents the (isotropic) hydrostatic pressure and the two other terms
describe flow--induced effects~\cite{Hess1975}. Specifically, $\mathbf{T}^{asy}$ includes asymmetric contributions, whereas
$\stl{\mathbf{T}}$ is a symmetric traceless tensor which can be written as 
\begin{align}
\label{Eq:Stress-Tensor}
\stl{\mathbf{T}} = 2 \eta_{iso} \mathbf{\Gamma} + \stl{\mathbf{T}_{al}}\, .
\end{align}
According to eq.~(\ref{Eq:Stress-Tensor}), $\stl{\mathbf{T}}$ splits into a {\em Newtonian} contribution (already present in fluids with vanishing orientational order) and a contribution depending explicitly on the 
$\mathbf{Q}$--tensor, that is, 
\begin{align}
\label{Eq:Stress-Alignment}
\stl{\mathbf{T}_{al}} = &\, \frac{\rho}{m} k_B T \left( - \sqrt{2} \frac{\tau_{qp}}{\tau_q} \mathbf{\Phi}' + \sqrt{2} \frac{\tau_{qp}}{\tau_q} \xi^2 \nabla^2 \mathbf{Q}\right)\nonumber \\ &+ 2\sigma \frac{\rho}{m} k_B T \left( \stl{\mathbf{Q} \cdot \mathbf{\Phi}'} - \xi^2 \stl{\mathbf{Q} \cdot \nabla^2 \mathbf{Q}} \right)\,.
\end{align}

In the present work we neglect the asymmetric part of the stress tensor ({\em i.e.}, $\mathbf{T}^{asy} = 0$), since it typically relaxes faster to zero than the relevant hydrodynamic processes considered. 
Further, in Newtonian flow regimes this antisymmetric stress is zero anyway~\cite{Hess2015,Heidenreich2009}.
\subsection{Explicit equations of motion}
\label{SubSec:02.2}
Equations~(\ref{Eq:Hess-Qtensor}) and~(\ref{Eq:Hess-Stensor}) can be rewritten in terms of scaled variables; this is described in detail in the Appendix (see also~\cite{LugoFrias2016}).
One obtains
\begin{align}
\label{Eq:Exp-Qtensor}
\frac{d \tilde{\mathbf{Q}}}{d \tilde{t}} &= \tilde{\xi}^2 \tilde{\nabla}^2 \tilde{\mathbf{Q}} + \tilde{\mathbf{H}}(\tilde{\mathbf{Q}},\tilde{\mathbf{v}}) \,,\\
\label{Eq:Exp-Stensor}
\frac{d \tilde{\mathbf{v}}}{d \tilde{t}} &= \frac{1}{\beta}\tilde{\nabla} \tilde{\mathbf{T}}\,,
\end{align}
where the tilde indicates scaled quantities and the parameter
$\beta$ appearing in eq.~(\ref{Eq:Exp-Stensor}) is defined as 
\begin{align}
\label{Eq:Beta}
\beta=\frac{24C^2\dot{\gamma} L^2 m}{B^2 k_B T \tau_q}\,.
\end{align}
This coefficient, or rather the ratio between $\beta$ and the scaled viscosity $\tilde{\eta}_{iso}$ defines the Reynolds number of the solvent 
\begin{align}
\label{Eq:Reynolds-Number}
Re = \frac{\beta}{\tilde{\eta}_{iso}} = \frac{\dot{\gamma} L^2 \rho}{{\eta}_{iso}}\,.
\end{align}
Experiments of shear--induced instabilities are typically performed at low Reynolds numbers, $Re\ll 1$~\cite{Chakraborty2010,Ganapathy2008}. 
In this limit the momentum balance equation (\ref{Eq:Exp-Stensor}) reduces to
\begin{align}
\label{Eq:Stensor_0}
\tilde{\nabla} \tilde{\mathbf{T}} = 0\,.
\end{align}
We note that due to the time dependence of $\tilde{\mathbf{Q}}(t)$, the total stress 
$ \tilde{\mathbf{T}}$ evaluated through eqs.~(\ref{Eq:Stress-Tensor}) and~(\ref{Eq:Stress-Alignment}) generally also depends
on time. However, at each time the total stress has to fulfill eq.~(\ref{Eq:Stensor_0}). 
The resulting velocity profile [obtained by solving eq.~(\ref{Eq:Stress-Tensor}) under the condition eq.~(\ref{Eq:Stensor_0})]
can therefore deviate from the linear profile assumed initially. This is clearly essential for the description of spatial symmetry--breaking such as shear banding.

In the following we drop the tilde $(\sim)$ on all variables; all quantities then appear with the same symbols as originally.  We also set $\sigma = 0$ since this parameter has minor effect on the dynamics of the system for planar Couette flow geometry (see~\cite{Hess2004,Hess2005}). 
Regarding the spatial variation of the ${\bf Q}$--tensor and the stress, we restrict ourselves to a one-dimensional investigation along the $y$--axis, {\em i.e.}, the direction of the shear gradient
(see fig.~\ref{Fig:Sketch}). Thus, we here exclude the possibility of banding in vorticity ($z$--) direction.

The resulting dynamical equations are simplified by expressing the $\mathbf{Q}$--tensor in terms of a standard orthonormal tensor basis, that is, 
$\mathbf{Q} = \sum_{i=0}^{4} q_i \mathbf{B}_i$~\cite{Kaiser1992}, where $q_0$, $q_1$ and $q_2$ are related to ordering within the $x$--$y$--plane ({\em i.e.}, the shear plane),
whereas $q_3$ and $q_4$ refer to out--of--plane ordering~\cite{Kaiser1992}. From the orthogonality of the basis functions 
($\mathbf{B}_i:\mathbf{B}_j = \delta_{ij}$) it follows that $q_i = \mathbf{Q}:\mathbf{B}_i$, which allows
to rewrite eqs.~(\ref{Eq:Exp-Qtensor}) into a set of scalar equations. Explicitly, one has
\begin{align}
\label{Eq:Scalar-Equations-Alignment}
\frac{d q_0}{d t} &= - \Phi_0 + \xi^2 \frac{\partial^2 q_0}{\partial y^2}\,, \nonumber \\
\frac{d q_1}{d t} &= - \Phi_1 + \dot{\gamma} \frac{\partial v}{\partial y} q_2 + \xi^2 \frac{\partial^2 q_1}{\partial y^2}\,, \nonumber \\
\frac{d q_2}{d t} &= - \Phi_2 - \dot{\gamma} \frac{\partial v}{\partial y} q_1  + \xi^2 \frac{\partial q_2}{\partial y^2} + \dot{\gamma} \lambda \frac{\partial v}{\partial y}\,, \\
\frac{d q_3}{d t} &= - \Phi_3 + \frac{1}{2} \dot{\gamma} \frac{\partial v}{\partial y} q_4  + \xi^2 \frac{\partial^2 q_3}{\partial y^2},, \nonumber \\
\frac{d q_4}{d t} &= - \Phi_4 - \frac{1}{2} \dot{\gamma} \frac{\partial v}{\partial y} q_3  + \xi^2 \frac{\partial^2 q_4}{\partial y^2}\,. \nonumber 
\end{align}
In eqs.~(\ref{Eq:Scalar-Equations-Alignment}) the quantities $\Phi_i$ are non--linear functions of the $q_i$ (stemming from the free--energy derivatives); they are given by
\begin{align}
\label{Eq:Scalar-Equations-DerEnergy}
\Phi_0 &= \left( \Theta - 3 q_0 + 2 q^2 \right) q_0 + 3 \left( q_1^{2} + q_2^{2} \right) - \frac{3}{2} \left( q_3^{2} - q_4^{2} \right)\,, \nonumber \\
\Phi_1 &= \left( \Theta + 6 q_0 + 2 q^2 \right) q_1 - \frac{3}{2} \sqrt{3} \left( q_3^{2} - q_4^{2} \right)\,, \nonumber \\
\Phi_2 &= \left( \Theta + 6 q_0 + 2 q^2 \right) q_2 - 3 \sqrt{3} q_3 q_4\,, \\
\Phi_3 &= \left( \Theta - 3 q_0 + 2 q^2 \right) q_3 - 3 \sqrt{3} \left( q_1 q_3 + q_2 q_4 \right)\,, \nonumber\\
\Phi_4 &= \left( \Theta - 3 q_0 + 2 q^2 \right) q_4 + 3 \sqrt{3} \left( q_1 q_4 - q_2 q_3 \right)\,,\nonumber
\end{align}
where $q^2 = \sum_{i=0}^4 q_i^2$. Finally, the momentum balance equation (\ref{Eq:Exp-Stensor})  becomes (in the regime of low Reynolds numbers)
\begin{align}
\label{Eq:Scalar-Equations-Stress}
0=\frac{\partial T_2}{\partial y} = \sqrt{2} \eta_{iso} \dot{\gamma} \frac{\partial^2 v}{\partial y^2} + \sqrt{2} \lambda \frac{\partial \Phi_2}{\partial y} - \sqrt{2} \lambda \xi^2 \frac{\partial^3 q_2}{\partial y^3}\,.
\end{align}
Equation~(\ref{Eq:Scalar-Equations-Stress}) indicates that the only non--vanishing component of the stress tensor is $T_2$,  which corresponds to the in--plane stress $T_{xy}$.

\subsection{Numerical calculations}
Equations~(\ref{Eq:Scalar-Equations-Alignment})--(\ref{Eq:Scalar-Equations-Stress}) are integrated numerically using a fourth order Runge--Kutta scheme~\cite{Press1996} with a fixed time step
$\Delta t = 2\times10^{-3}$ and a grid spacing of $\Delta y = 5\times10^{-3}$.
The gradient terms are discretized by a central finite difference scheme of fourth order. At the boundaries, asymmetric stencils (using only available grid points) are implemented~\cite{Press1996}. 
The calculations are initialized with values of $q_0, \cdots, q_4$ matching the boundary conditions (see below), to accelerate the calculations we
additionally use a small random perturbation. To find steady configurations of the system we monitor the evolution of $q_0, \cdots, q_4$ and $T_2$, disregarding transient behavior. 
The resulting dynamical states are characterized employing an algorithm that recognizes the periodicity and sign change of the time dependent components $q_0, \cdots, q_4$~\cite{Rienacker2000,Hess1994}.

It turns out that the solution of eqs.~(\ref{Eq:Scalar-Equations-Alignment})--(\ref{Eq:Scalar-Equations-Stress}) is quite sensitive to initial conditions; thus all
calculations have been repeated several times. Further, we have checked that the steady--state solutions do not change with decreasing $\Delta t$ (however, the numerical stability does depend on the grid spacing).

Regarding the boundary conditions at the plates ($y=\pm L$), we assume "strong anchoring" conditions, that is, the $\mathbf{Q}$--tensor at the plates 
is constant in time, but may have different symmetries. We further assume that the degree of ordering is given by the corresponding equilibrium value. 

\begin{figure}
\centering
\includegraphics[height=4.5cm,keepaspectratio]{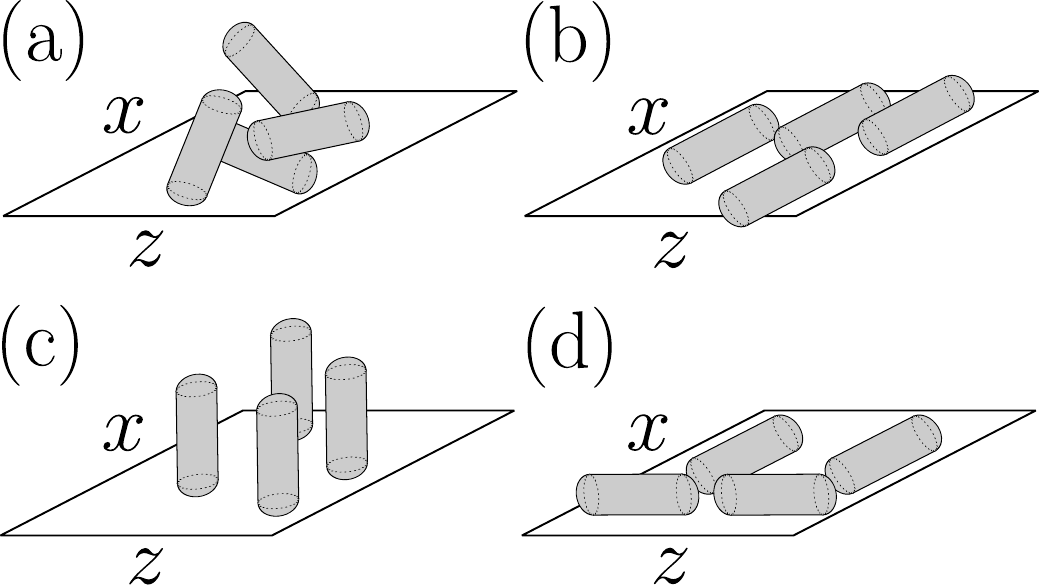}
\caption{Sketch of the boundary conditions applied to the tensor $\mathbf{Q}$ at the plates: (a)~Isotropic [see eq.~(\ref{Eq:Bound-Isotropic})] (b)~Planar nematic [eq.~(\ref{Eq:Bound-PlanNematic})] (c)~Vertical nematic [eq.~(\ref{Eq:Bound-VertNematic})] and (d)~Planar degenerate [eq.~(\ref{Eq:Bound-PlanDegenerate})].}
\label{Fig:Anchoring}
\end{figure}

Though these assumptions are clearly an idealization, we note that, from an experimental point of view, it is indeed possible to realize different boundary conditions
by chemical or mechanical treatment of the plates~\cite{Sheng1982,Jerome1991,Ruths1996,Manyuhina2010}. Here we focus on the following cases (see fig.~\ref{Fig:Anchoring}).
\begin{align}
\label{Eq:Bound-Isotropic}
&\mathbf{Q}\Big|^{y=L}_{y=-L} = 0 & &\text{(Isotropic)}\,,\\
\label{Eq:Bound-PlanNematic}
&\mathbf{Q}\Big|^{y=L}_{y=-L} = \sqrt{\frac{3}{2}}\,{\mu}^{eq}_3\stl{\hat{\mathbf{e}}_x \hat{\mathbf{e}}_x}  & &\text{(Planar nematic)}\,,\\
\label{Eq:Bound-VertNematic}
&\mathbf{Q}\Big|^{y=L}_{y=-L} = \sqrt{\frac{3}{2}}\,{\mu}^{eq}_3\stl{\hat{\mathbf{e}}_y \hat{\mathbf{e}}_y}  & &\text{(Vertical nematic)}\,,\\
\label{Eq:Bound-PlanDegenerate}
&\mathbf{Q}\Big|^{y=L}_{y=-L} = -\sqrt{\frac{3}{2}}\,{\mu}^{eq}_3\stl{\hat{\mathbf{e}}_y \hat{\mathbf{e}}_y}  & & \text{(Planar degenerate)}\,,
\end{align}
where ${\mu}^{eq}_3$ is the equilibrium value of the alignment tensor in the nematic phase. 
Equations~(\ref{Eq:Bound-Isotropic}) and~(\ref{Eq:Bound-PlanDegenerate}) describe disordered states where the rod orientations are distributed, either in all three directions ("isotropic")
or within the plane of the plates, {\em i.e.}, in the $x$--$z$ plane ("planar degenerate"). The other two boundary conditions correspond to nematic states, with the director lying either in the
plane of the plates ("planar nematic") or along the gradient ($y$--) direction ("vertical nematic"). 
We note that the latter boundary condition is sometimes referred to as "homeotropic".
Regarding the velocity field, we implement no--slip boundary conditions, that is $v_x(y=\pm1)=\pm 1$ (in reduced units).

\section{Results and discussion}
\label{Sec:03}
In this section we present numerical results for shear--driven systems whose equilibrium configuration ($\dot{\gamma}=0$) is either isotropic [characterized by $\Theta>9/8$ in the scaled free energy~(\ref{Eq:Scaled-DevEnergy})] or nematic [$\Theta <0$].
We divide the discussion into two parts. In the first part (sect.~\ref{SubSec:03.1}) we focus on the spatially homogeneous system. Here we explore how the non--zero component
of the stress tensor, $T_2$, is affected by the temporal evolution of $\mathbf{Q}(\mathbf{r},t)$ and use this information to predict the formation of spatial instabilities. 
The second part (sect.~\ref{SubSec:03.2}) is devoted 
to the spatial--temporal behavior for initially isotropic and nematic systems, as well as to the impact of boundary conditions.

\subsection{Homogeneous solutions}
\label{SubSec:03.1}
Here we consider spatially homogeneous systems where the boundaries do not play a role, corresponding to the limit of infinite plate separation, {\em i.e.}, $L \rightarrow \infty$.
The scaled correlation length appearing in eq.~(\ref{Eq:Exp-Qtensor})
becomes zero and thus, all gradient terms in eqs.~(\ref{Eq:Scalar-Equations-Alignment}) and~(\ref{Eq:Scalar-Equations-Stress}) vanish. In particular, the stress $T_2$ takes the form
\begin{align}
\label{Scalar-Equations-Stress-Hom}
T_2(t) = \sqrt{2} \eta_{iso} \dot{\gamma} + \sqrt{2} \lambda \Phi_2\,.
\end{align}
Here we set $\eta_{iso}=1.0$.
\subsubsection{Homogeneous systems sheared from the isotropic state }
\label{SubSubSec:03.1.1}
\begin{figure}
\centering
\includegraphics[height=4.50cm,keepaspectratio]{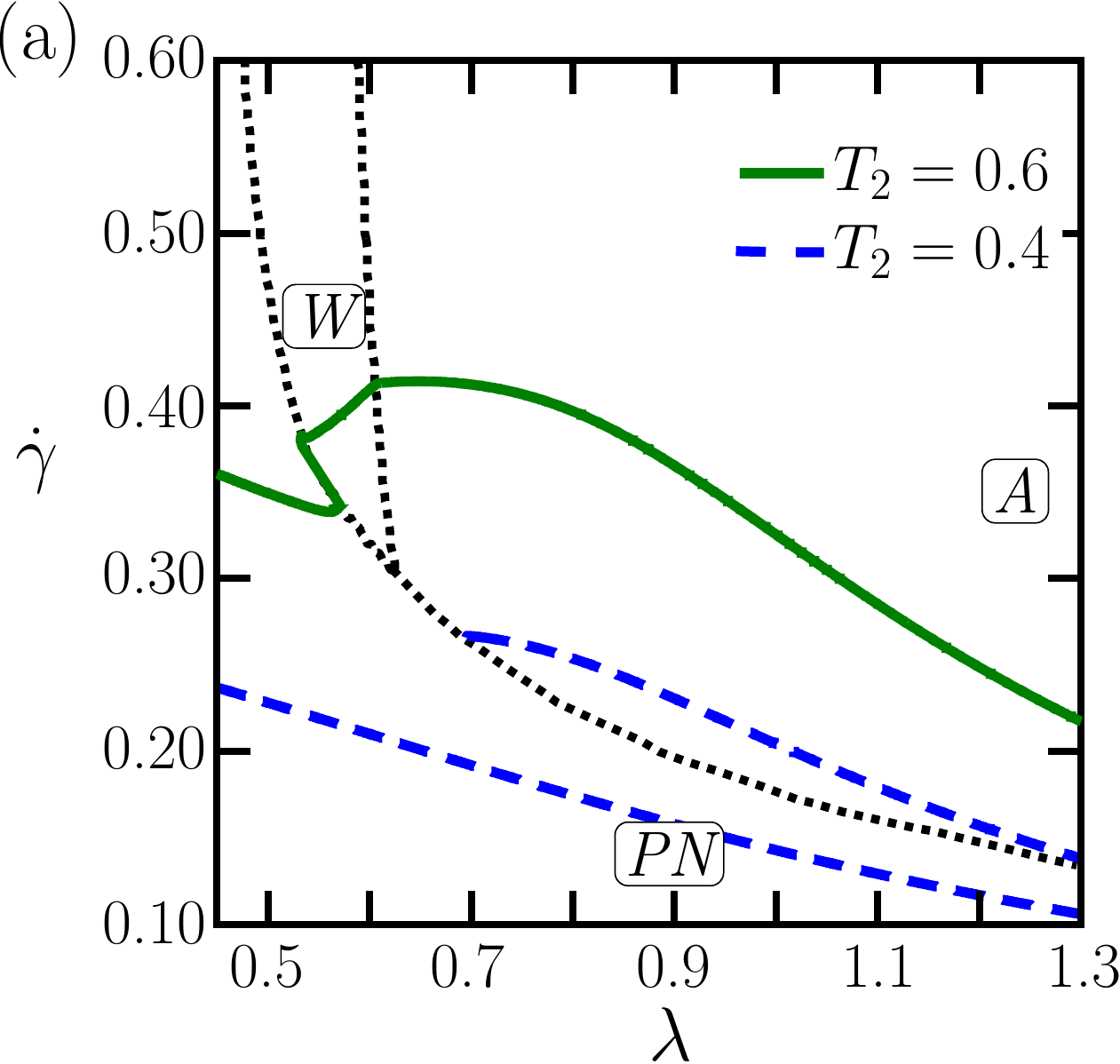}
\\\vspace{2mm}
\includegraphics[width=4cm,keepaspectratio]{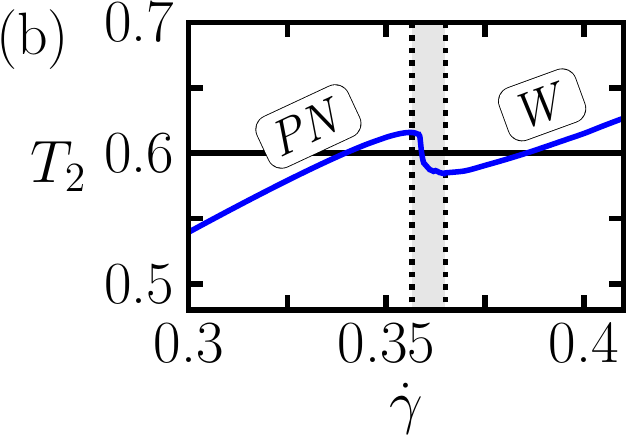}$\quad$
\includegraphics[width=4cm,keepaspectratio]{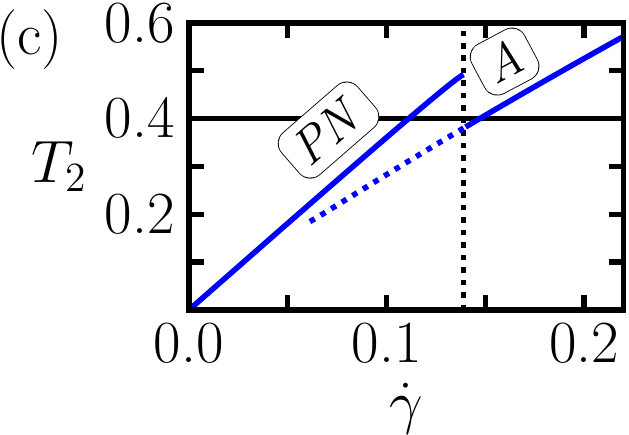}
\caption{(Color online) (a)~State diagram in the plane spanned by tumbling parameter ($\lambda$) and shear rate ($\dot\gamma$) at $\Theta = 1.20$
(isotropic equilibrium system). The dotted gray lines indicate the boundaries between three different states: paranematic (PN), shear--aligned (A) and wagging (W). The dashed (blue) and solid (green) lines connect points with constant stress $T_2 = 0.4$ and $T_2 = 0.6$, respectively. (Second row) Homogeneous flow curves $T_2(\dot\gamma)$ for (b)~$\lambda = 0.55$ and (c)~$\lambda = 1.25$.}
\label{Hom-Isotropic}
\end{figure}

We start by considering systems whose equilibrium state is isotropic ($\Theta = 1.20$). Increasing the shear rate from zero the system 
first develops a paranematic (PN), steady state characterized by weak (yet non--zero) nematic order; this is
illustrated in fig.~\ref{Hom-Isotropic}(a). 
The behavior upon further increase of $\dot\gamma$ then depends
on the tumbling parameter $\lambda$ (recall that the latter is a measure of the aspect ratio). 
For $\lambda\gtrsim 0.62$ one observes a transition from paranematic to shear--aligned (A) state; 
the latter is also characterized by a 
time--independent director (as is the PN state), but the degree of ordering which can be quantified, e.g., by the norm $||\mathbf{Q}|| = \sum_{i=0}^4 q_i^2$, is larger~\cite{Pardowitz1980,Kaiser1992}. 
The PN--A transition is accompanied by a narrow region of bistability (not visible in fig.~\ref{Hom-Isotropic}); in this regime the system's degree of ordering depends on the initial condition.
This feature is reminiscent of the first--order isotropic--nematic transition in equilibrium. For smaller values of the tumbling parameter ($\lambda\lesssim 0.62$)
the $A$ state is unstable; here the system develops wagging (W) motion characterized by regular oscillations of the nematic director within the shear plane.
Overall, the behavior found in the present calculations agrees qualitatively with that reported in ref.~\cite{Strehober2013}; the quantitative data for the boundary lines somewhat differ due to 
the different scaling of the free energy (see Appendix).

For each of the parameter sets ($\dot\gamma,\lambda)$ we have calculated the stress $T_2$ (in the oscillatory W state, we have averaged $T_2(t)$ over one period of time). Importantly, it turns out
that different parameter sets can lead to the same value of $T_2$. To illustrate this point, fig.~\ref{Hom-Isotropic}(a) includes dashed (blue) lines and solid (green) lines corresponding to two constant values of
$T_2$. Moreover, there are several regions of $\lambda$ where $T_2$ assumes the same value for different shear rates. For example, at $\lambda=0.55$ there are three values
of $\dot\gamma$ with
$T_2=0.6$, and for $\lambda\geq 0.7$ one finds two solutions with $T_2=0.4$. 

Given this multivalued behavior, it is interesting to consider corresponding flow curves $T_2(\dot\gamma)$. Results for $\lambda= 0.55$ and $\lambda =1.25$
are plotted in figs.~(\ref{Hom-Isotropic}(b) and~(\ref{Hom-Isotropic}(c), respectively.  The flow curve for $\lambda = 0.55$ [see fig.~\ref{Hom-Isotropic}(b)] displays a region with a negative slope ($dT_2/d\dot\gamma<0$) between $\dot\gamma \approx 0.357$ and $\dot\gamma \approx 0.365$. 
Within this region the homogeneous flow is mechanically unstable, and as one might expect (and will be explicitly shown in sect.~\ref{SubSubSec:03.2.1}), the system forms
a spatially inhomogeneous, shear banded state. We also note that the shear rate $\dot\gamma\approx 0.359$ characterized by $T_2=0.6$ and $dT_2/d\dot\gamma<0$ agrees roughly with
the corresponding point on the boundary line PN--W in fig.~\ref{Hom-Isotropic}(a). This indicates that the orientational transition from the (steady) PN state to the (oscillatory) W state, on the one side, and
the shear banding instability, on the other side, are closely intercorrelated.

In contrast, for $\lambda = 1.25$ [see fig.~\ref{Hom-Isotropic}(c)] the flow curve does not display a region with negative slope. Rather one observes a 
discontinuity and, associated with this, hysteretic behavior. Upon increase of $\dot\gamma$ from the small values, {\em i.e.}, from the paranematic (PN) state, the systems discontinuously 
"jumps" to the aligned (A) state only at $\dot\gamma \approx 0.14$ [which is above the upper blue dashed line  in fig.~\ref{Hom-Isotropic}(a)]. However, decreasing $\dot\gamma$ starting from the large shear rates characterizing the A state, the jump occurs at the much smaller shear rate $\dot\gamma\approx  0.06$. 
As we will show in sect.~\ref{SubSubSec:03.2.2}, the formation of shear bands in this case ($\lambda=1.25$) strongly depends on the boundary conditions.

\subsubsection{Homogeneous systems sheared from the nematic state}
\label{SubSubSec:03.1.2}
\begin{figure}
\centering
\includegraphics[height=4.70cm,keepaspectratio]{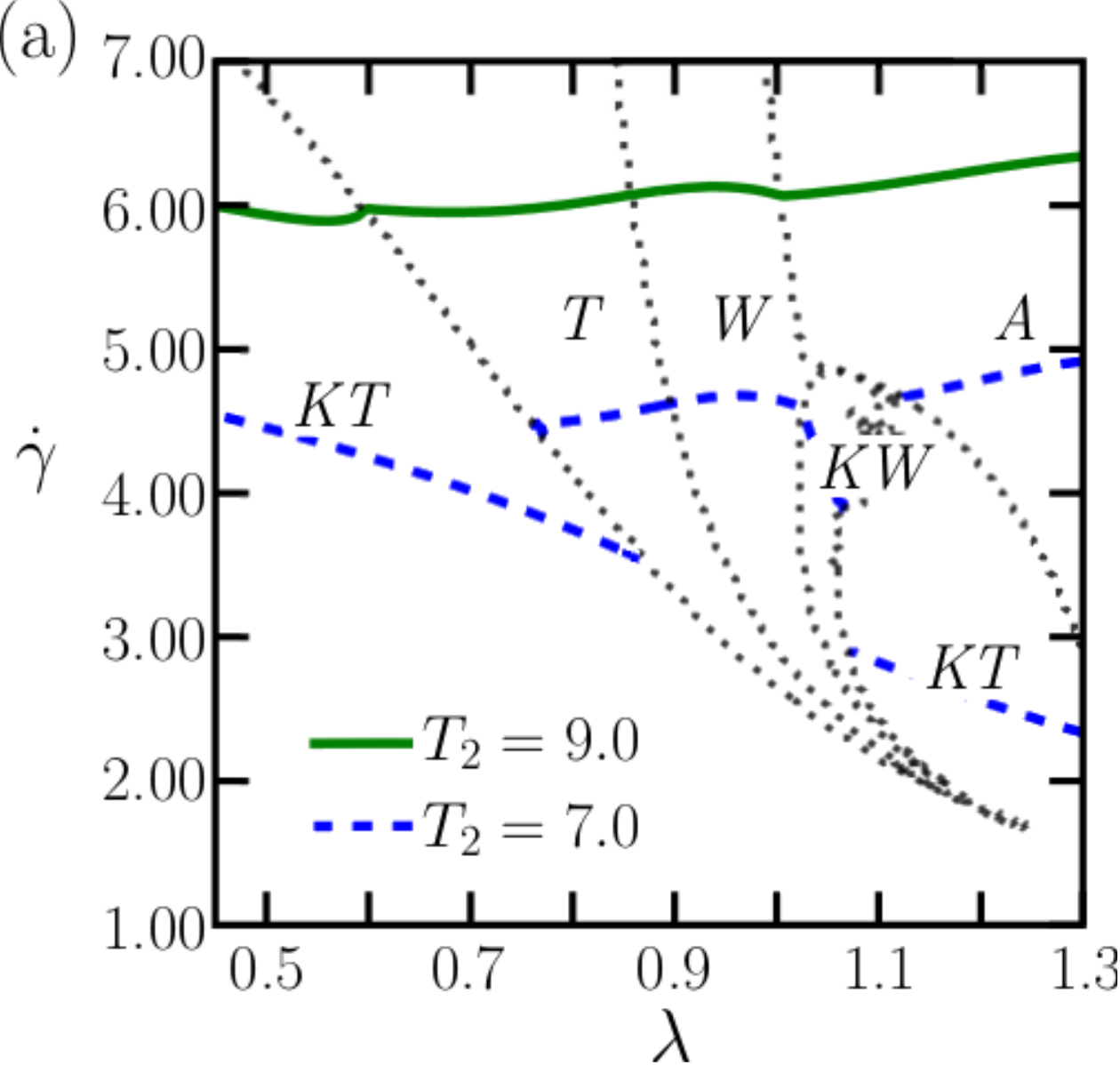}
\\\vspace{2mm}
\includegraphics[width=3.65cm,keepaspectratio]{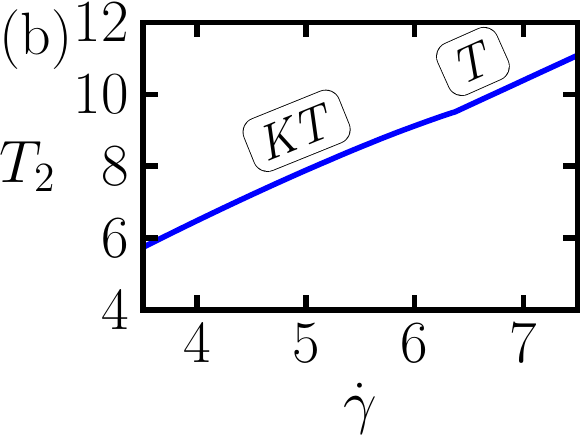}$\quad$
\includegraphics[width=3.65cm,keepaspectratio]{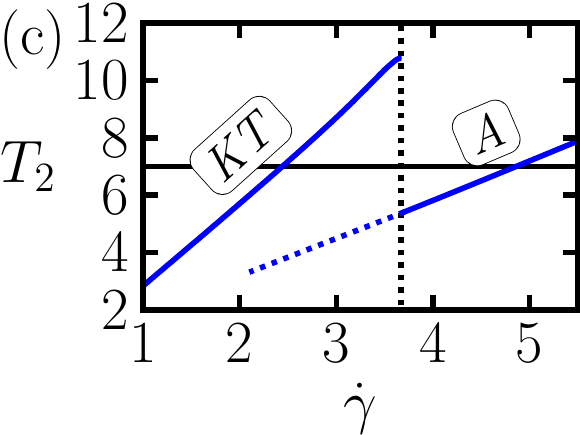}
\caption{(Color online) (a)~State diagram at $\Theta = -0.25$. The dotted gray lines indicate the boundaries between 
the different dynamical states: wagging (W), tumbling (T), kayaking--tumbling (KT), kayaking--wagging (KW), and shear--alignment (A).
The dashed (blue) and solid (green) lines connect points with constant stress $T_2 = 7.0$ and $T_2 = 9.0$, respectively. (Second row) Homogeneous flow curves $T_2(\dot\gamma)$ for (b)~$\lambda = 0.55$ and (c)~$\lambda = 1.25$.}
\label{Hom-Nematic}
\end{figure}

We now turn to systems which, in thermal equilibrium, are deep within the nematic phase. Specifically, we set $\Theta = -0.25$ in eq.~(\ref{Eq:Scalar-Equations-DerEnergy}). 
Similar to previous work~\cite{Rienacker2002,Strehober2013}  we find that shear can induce a variety of time--dependent dynamical states [in addition to the W motion already appearing in initially isotropic systems, see fig.~\ref{Hom-Isotropic}(a)], as well as
a shear--aligned (A) steady state. An overview is given in fig.~\ref{Hom-Nematic}(a).

In the wagging and tumbling (T) state, the nematic director performs regular, oscillatory motion within the shear plane ($q_4(t)=q_5(t)=0\,\forall t$), whereas it displays out--of--plane (yet regular) oscillations
in the kayak--tumbling (KT) and kayak--wagging (KW) state ($q_i(t)\neq 0$ $\forall\, i=0,\ldots,4$). Only in the A state the director stays constant in time. 
Note that, contrary to the case considered before (see fig.~\ref{Hom-Isotropic}), there is no
paranematic (PN) state at $\Theta=-0.25$ since the system is deep in the nematic regime. We also note that  earlier studies~\cite{Rienacker2002,Strehober2013} investigating 
similar values of $\Theta$ have reported the occurrence of a region characterized by irregular and even chaotic motion of the director. 
This region is located around the point where the KT, KW and A states meet. Here we did not detect such a region because our algorithm does not resolve Lyapunov exponents.

We now turn to the resulting shear stress. The dashed (blue) and solid (green) lines in fig.~\ref{Hom-Nematic}(a) indicate parameter sets at which the orientational dynamics yield the constant stress--values $T_2=9.0$ and $T_2=7.0$, respectively.
In the first case, the line provides a unique relation in the sense that
an increase of $\dot\gamma$ at fixed $\lambda$ yields only one crossing with this line. This is different for the case $T_2=7.0$ where, depending on $\lambda$, one or two crossings can be observed.
Exemplary flow curves for two values of the tumbling parameter are shown in the bottom parts of fig.~\ref{Hom-Nematic}. At $\lambda \approx 0.55$ [see fig.~\ref{Hom-Nematic}(b)], where each of the constant--pressure lines is crossed only once, one observes a monotonic increase of $T_2$ with $\dot\gamma$. In particular, there is no discontinuity or cusp even 
at $\dot\gamma\approx 6.5$, where the underlying orientational dynamics changes from out--of--plane kayaking--tumbling to in--plane tumbling.
We note, however, that a systematic bifurcation analysis (such as the one in ref.~\cite{Strehober2013}, where a very similar system was considered) would presumably reveal 
a {\em bistable} region characterized by the presence of (at least) two attractors between the pure KT and the pure T state.
In such a situation, the pressure $T_2$ would not be uniquely defined. This aspect certainly deserves more attention in a future study.

We now consider the case $\lambda \approx 1.25$ in fig.~\ref{Hom-Nematic}(a), where an increase of $\dot\gamma$ from small values yields two crossings with the constant--pressure curve
$T_2=7.0$. As seen from fig.~\ref{Hom-Nematic}(c), the flow curve exhibits a pronounced discontinuity related to the transformation
of the (out--of--plane) oscillating KT state into the shear--aligned (A) steady state at $\dot{\gamma} \approx 3.67$. One also recognizes a strong dependence on initial conditions (hysteresis), similar to 
the situation discussed in fig.~\ref{Hom-Isotropic}(c). As we will later discuss in sect.~\ref{SubSubSec:03.2.2}, the initially nematic system at $\lambda \approx 1.25$ indeed
displays shear banding.
\begin{figure*}
\centering
\includegraphics[height=4.75cm,keepaspectratio]{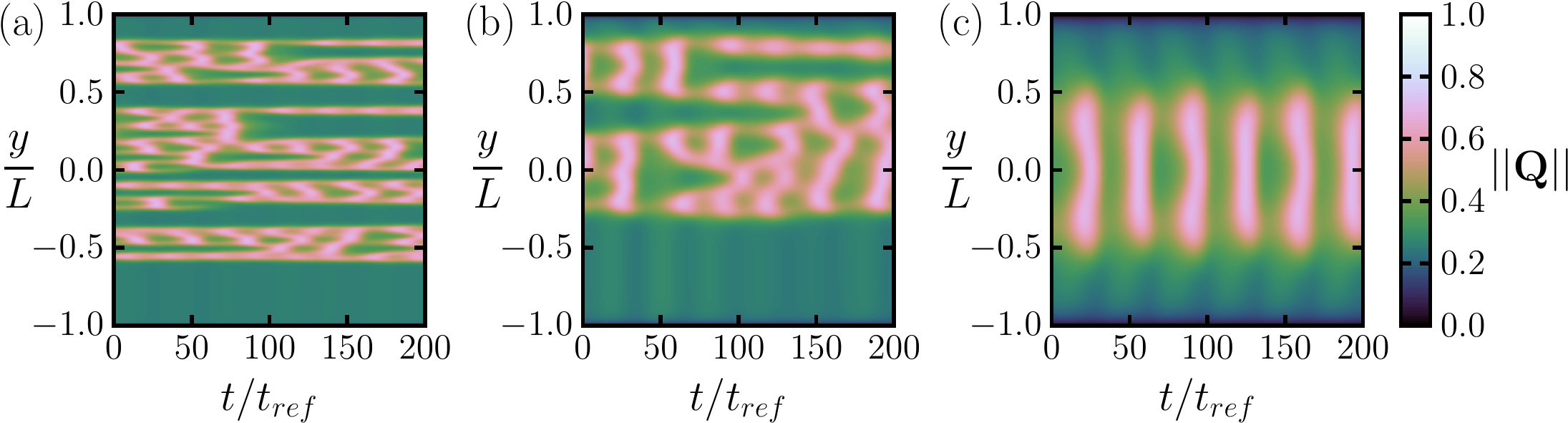}
\caption{(Color online) Space--time plot of the norm of the $\mathbf{Q}$--tensor at $\dot{\gamma}=0.365$, tumbling parameter $\lambda=0.55$ and different correlation lengths (a)~$\xi^2  =10^{-5}$, (b)~$\xi^2 = 10^{-4}$ and (c)~$\xi^2 = 10^{-3}$. The equilibrium state is isotropic ($\Theta=1.20$).}
\label{Inhom-Isotropic}
\end{figure*}

\subsection{Spatiotemporal behavior and shear banding}
\label{SubSec:03.2}
In the preceding discussion we have found indications of the formation of inhomogeneous states in both, systems sheared from the isotropic and
systems sheared from the nematic phase. We now analyze the corresponding systems (characterized by certain values of the tumbling parameter) 
further by calculating the full, spatiotemporal behavior of the $\mathbf{Q}$--tensor and the resulting shear stress $T_2$.
To this end we have solved numerically eqs.~(\ref{Eq:Scalar-Equations-Alignment})--(\ref{Eq:Scalar-Equations-Stress}) using the methodology described at the end of sect.~\ref{Sec:02}. 

Our discussion in this section is divided into two parts, covering the role of the two key factors impacting the spatial structure of the inhomogeneous systems. 
These are, first, the correlation length $\xi$, which appears as a prefactor of the gradient term in the orientational free energy density [see eq.~(\ref{Eq:Elastic-Energy})],
and second, the boundary condition for $\mathbf{Q}$ at the plates [see eqs.~(\ref{Eq:Bound-Isotropic})--(\ref{Eq:Bound-PlanDegenerate})].
The impact of $\xi$ is discussed in sect.~\ref{SubSubSec:03.2.1}, where we fix the boundary conditions according to the equilibrium configuration of the system.
In sect.~\ref{SubSubSec:03.2.2} we then explore the role of different boundary conditions.

\subsubsection{Impact of the correlation length}
\label{SubSubSec:03.2.1}

\paragraph{Initially isotropic system}
We first consider the system at $\Theta=1.20$ and $\lambda=0.55$, where we have observed a region of negative slope in the corresponding flow curve, $T_2(\dot\gamma)$
[see fig.~\ref{Hom-Isotropic}(b)]. Here we focus on a shear rate within this regime, $\dot\gamma=0.365$. 
In figs.~\ref{Inhom-Isotropic}(a)--(c) we show the space--time evolution of the norm of the $\mathbf{Q}$--tensor at three values of the correlation length.  
Because the equilibrium state is isotropic, we choose the boundary conditions according to eq.~(\ref{Eq:Bound-Isotropic}), that is, the boundaries do not support any 
orientational ordering.
\begin{figure}
\centering
\includegraphics[height=4.0cm,width=3.2cm]{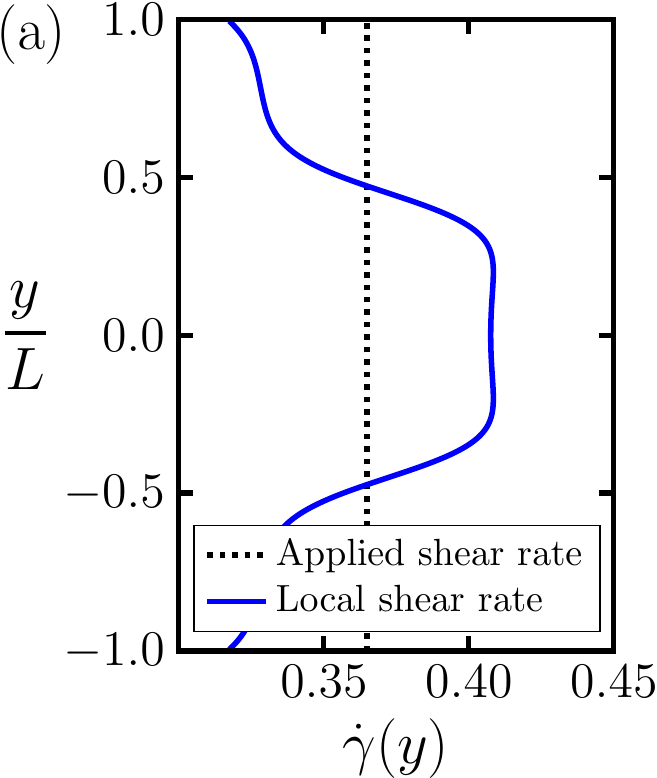}
\includegraphics[height=4.0cm,keepaspectratio]{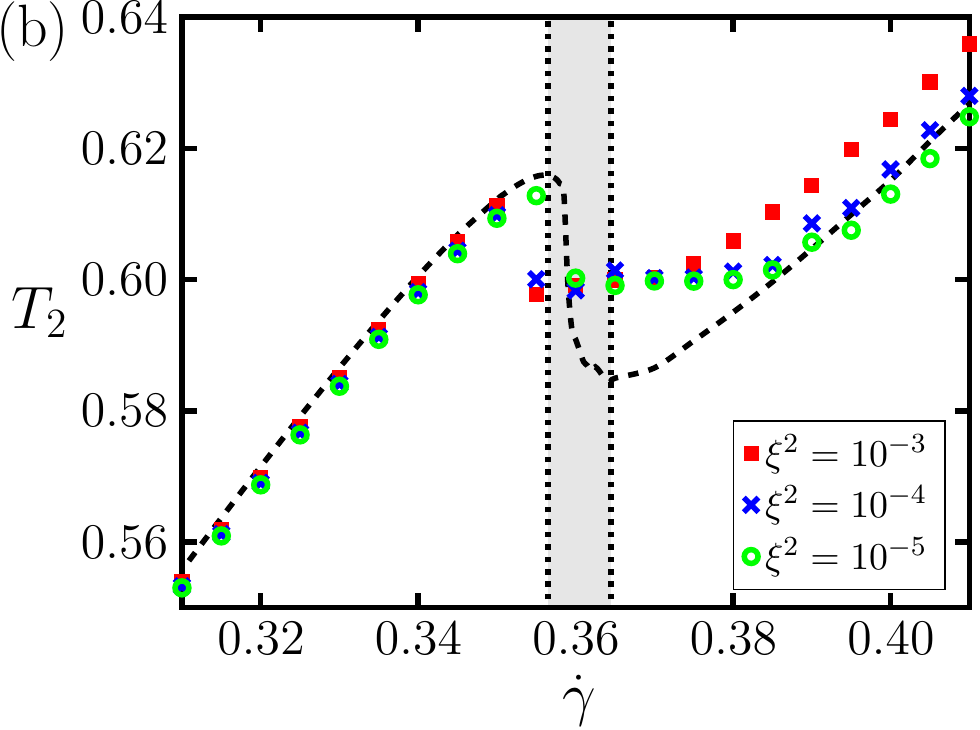}
\caption{(Color online) (a) Local shear rate within the banded state of the initially isotropic system ($\Theta=1.20$, $\lambda=0.55$, average (applied) 
shear rate $\dot\gamma=0.365$).
(b) Inhomogeneous flow curves at different correlation lengths. The symbols $\blacksquare$ (red), $\times$ (blue) and $\circ$ (green) correspond to $\xi^2 = 10^{-5}$, $\xi^2 = 10^{-4}$ and $\xi^2 = 10^{-3}$, respectively. As a reference the homogeneous flow curve is included (black dashed line).}
\label{Inhom-Flow-Isotropic}
\end{figure}

Still, as seen from fig.~\ref{Inhom-Isotropic}(a), the system forms spatiotemporal structures with locally large values of $||\mathbf{Q}||$ 
already at the smallest correlation length considered, $\xi = 10^{-5}$. Here, the observed pattern is rather "loose" with its width
changing in time. We also find that, within the inhomogeneous regions, $||\mathbf{Q}||$ is oscillating in time. 
A closer analysis reveals that the oscillations are consistent with a wagging (W) state. Outside the inhomogeneous regions, $||\mathbf{Q}||$ takes values typical of a paranematic (PN) state. This
behavior is, to some extent, expected since the value of $\dot\gamma$ considered in fig.~\ref{Inhom-Isotropic} is very close to the boundary between the PN and W
state (see fig.~\ref{Hom-Isotropic}). 

Upon increasing the correlation length, $\xi$, we observe from figs.~\ref{Inhom-Isotropic}(b) and~\ref{Inhom-Isotropic}(c) that the regions characterized by W motion become more defined, both in terms
of the shape of the emerging shear band, and in terms of the (increasingly regular) oscillatory motion of the order parameter. At the same time the interface between the W and PN region
becomes broader. As a consequence, the W oscillations are transferred to some extent into the outer region, however, with a very small amplitude.
 
A further illustration of the presence of shear bands is plotted in fig.~\ref{Inhom-Flow-Isotropic}(a), where we present the {\em local} shear rate, $\dot\gamma(y)$, 
for the system at $\xi^2=10^{-3}$.  
It is seen that the band in the middle of the system, where the orientational dynamics is of W type, is characterized by a significantly higher shear rate than the PN state close to the boundaries.
To complete the picture, fig.~\ref{Inhom-Flow-Isotropic}(b) shows flow curves obtained for the inhomogeneous (initially isotropic) systems at different correlation length.
Following previous studies~\cite{Radulescu2000} we have obtained these curves by calculating the mean value of $T_2$ increasing gradually from lower to larger values of $\dot{\gamma}$.
As a reference, the corresponding homogeneous flow curve [see fig.~\ref{Hom-Isotropic}(b)] is included. Interestingly, the value of $T_2$ corresponding to the banded state is essentially {\em independent}
of $\xi$; in other words, the value of $T_2$ 
appears to be unique. This observation is consistent with previous studies on the basis of both, a $\mathbf{Q}$--tensor model~\cite{Olmsted1999} and the DJS model~\cite{Olmsted2000}.
We further observe from fig.~\ref{Inhom-Flow-Isotropic}(b) that there is a slight dependence on the value of $T_2$ on $\xi$ at high shear rates beyond the banded state. 
This is an effect stemming from the inhomogeneities induced by the confining walls: the larger $\xi$, the larger is the extent of these inhomogeneities into the bulk--like region between the plates.
In fact, by excluding these regions from the calculations, the results for $T_2$ completely agree for different $\xi$. 
\begin{figure*}
\centering
\includegraphics[height=4.75cm,keepaspectratio]{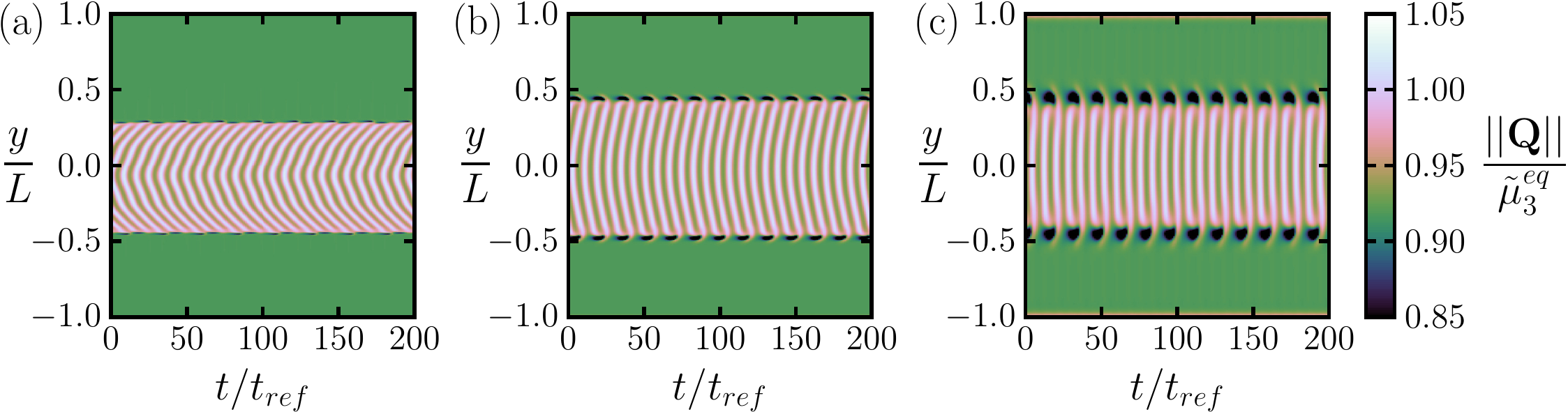}
\caption{(Color online) Space--time plot of the norm of the $\mathbf{Q}$--tensor at $\dot{\gamma}=3.65$, $\lambda=1.25$ and different correlation lengths (a)~$\xi^2  =10^{-5}$, (b)~$\xi^2 = 10^{-4}$ and (c)~$\xi^2 = 10^{-3}$. The equilibrium state $(\dot{\gamma} = 0)$ is nematic ($\Theta=-0.25$).}
\label{Inhom-Nematic}
\end{figure*}
 
So far we have focused on an initially isotropic system at $\lambda=0.55$. At the larger tumbling parameter $\lambda = 1.25$,
where the homogeneous calculations [see fig.~\ref{Hom-Isotropic}(c)] yield a discontinuous flow curve $T_2(\dot\gamma)$ (rather than one with negative slope), 
the results from the spatially--resolved calculations are more complex. For isotropic boundary conditions we did not find 
shear banding behavior, regardless of the value of $\xi$. However, using boundary conditions which support nematic ordering (such as the planar alignment in eq.~(\ref{Eq:Bound-VertNematic})
or the vertical alignment in eq.~(\ref{Eq:Bound-PlanDegenerate}))
we do find a well--defined shear band. This point will be further discussed in sect.~\ref{SubSubSec:03.2.2}.

\paragraph{Initially nematic system}
We now turn to the system at $\Theta = -0.25$ and $\lambda = 1.25$, where the homogeneous flow curve [see fig.~\ref{Hom-Nematic}(c)] is discontinuous.
Results for the norm of $\mathbf{Q}$ as function of space and time are shown in fig.~\ref{Eq:Hess-Stensor}, where we assumed equivalent planar nematic boundary conditions
[see eq.~(\ref{Eq:Bound-PlanNematic})], but different correlation lengths.

In all cases, one observes a clear spatial separation of the system into an inner band, where the orientational behavior corresponds to the kayak--tumbling (KT) state, and an outer region
where the system is in a shear--aligned (A) state. Upon increase of $\xi$ the width of the KT band widens, while the oscillations within the band become more and more regular.

\begin{figure}
\centering
\includegraphics[height=4.0cm,width=3.2cm]{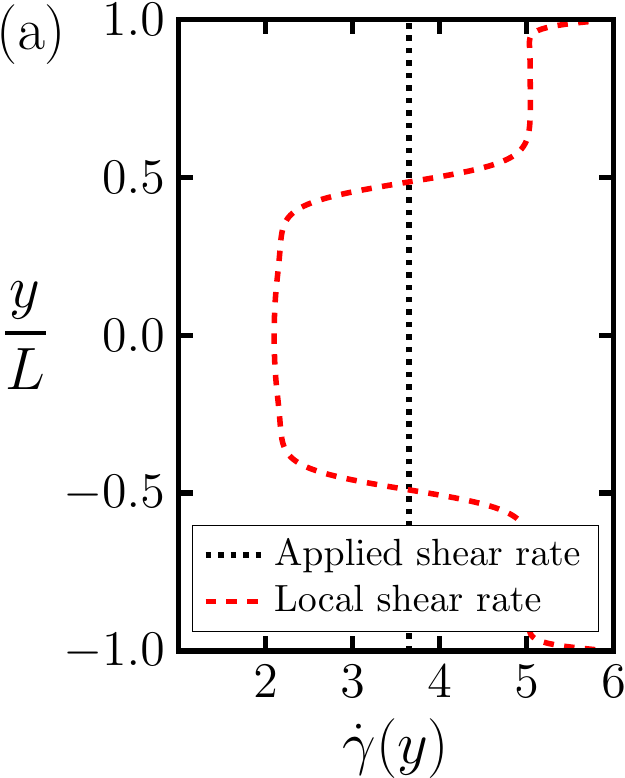}
\includegraphics[height=4.0cm,keepaspectratio]{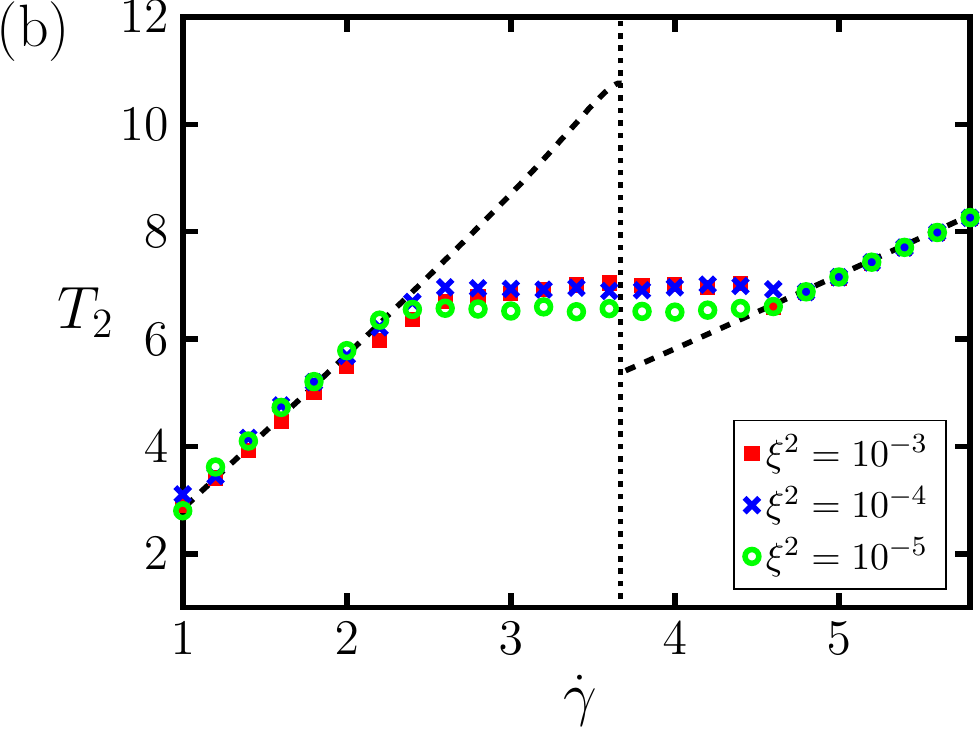}
\caption{(Color online) (a) Local shear rate within the banded state of the initially nematic system ($\Theta=-0.25$, $\lambda=1.25$, average (applied) shear rate $\dot\gamma=3.65$).
(b) Inhomogeneous flow curves at different correlation lengths. The symbols $\blacksquare$ (red), $\times$ (blue) and $\circ$ (green) correspond to $\xi^2 = 10^{-5}$, $\xi^2 = 10^{-4}$ and $\xi^2 = 10^{-3}$, respectively. As a reference the homogeneous flow curve is included (black dashed line).}
\label{Inhom-Flow-Nematic}
\end{figure}

Corresponding results for the local shear rate and the inhomogeneous flow curves are given in fig.~\ref{Inhom-Flow-Nematic}.
Compared to the initially isotropic system, we see from fig.~\ref{Inhom-Flow-Nematic}(a) that the oscillatory (KT) band is characterized by a {\em larger} shear rate than the regions close to the boundaries.
A further difference comes up when we consider in fig.~\ref{Inhom-Flow-Nematic}(b) the values of the stress plateau in the inhomogeneous flow curve. Here we find a dependence on the correlation length; that is,
the value of $T_2$ at the plateau increases with $\xi$. This contrasts with our corresponding results for the initially isotropic system (see the discussion of fig.~\ref{Inhom-Flow-Isotropic}).
To which extent this dependency is subject to the initial conditions of the numerical calculations is a point which has remained elusive so far.
\subsubsection{Role of the boundary conditions}
\label{SubSubSec:03.2.2}
This final section is devoted to the role of the boundary conditions [see eqs.~(\ref{Eq:Bound-Isotropic})--(\ref{Eq:Bound-PlanDegenerate})], which we here assume to be freely selectable irrespective of the initial state of the equilibrium system. 
The correlation length is set to a constant value of $\xi^2 = 10^{-5}$ (higher values yield very similar results).

\begin{figure}
\centering
\includegraphics[height=4.70cm,keepaspectratio]{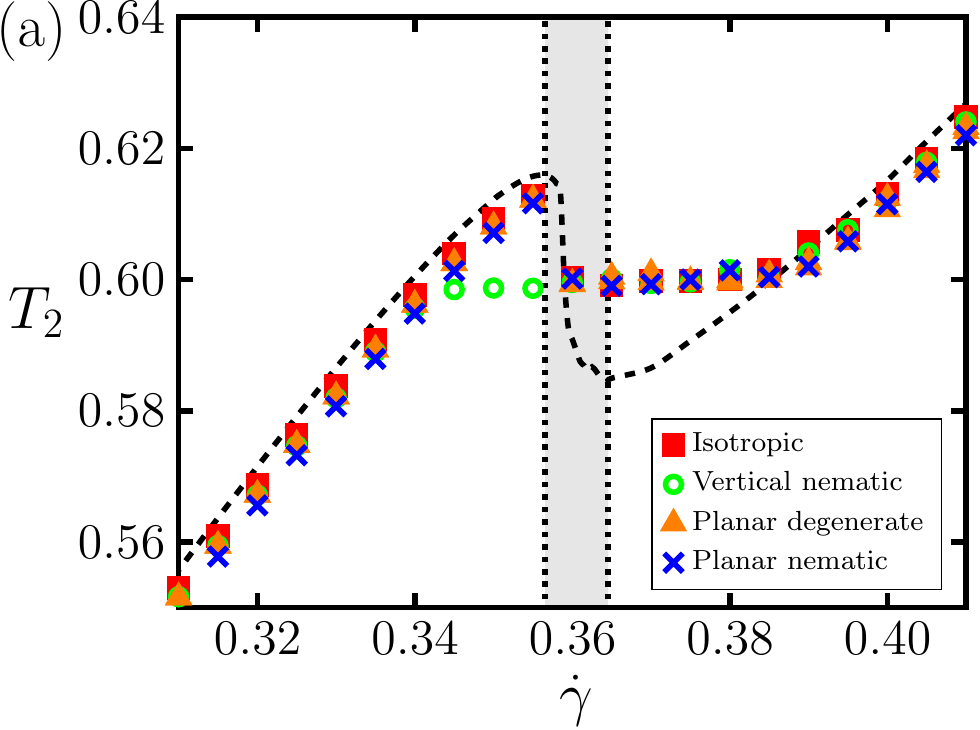}
\includegraphics[height=4.70cm,keepaspectratio]{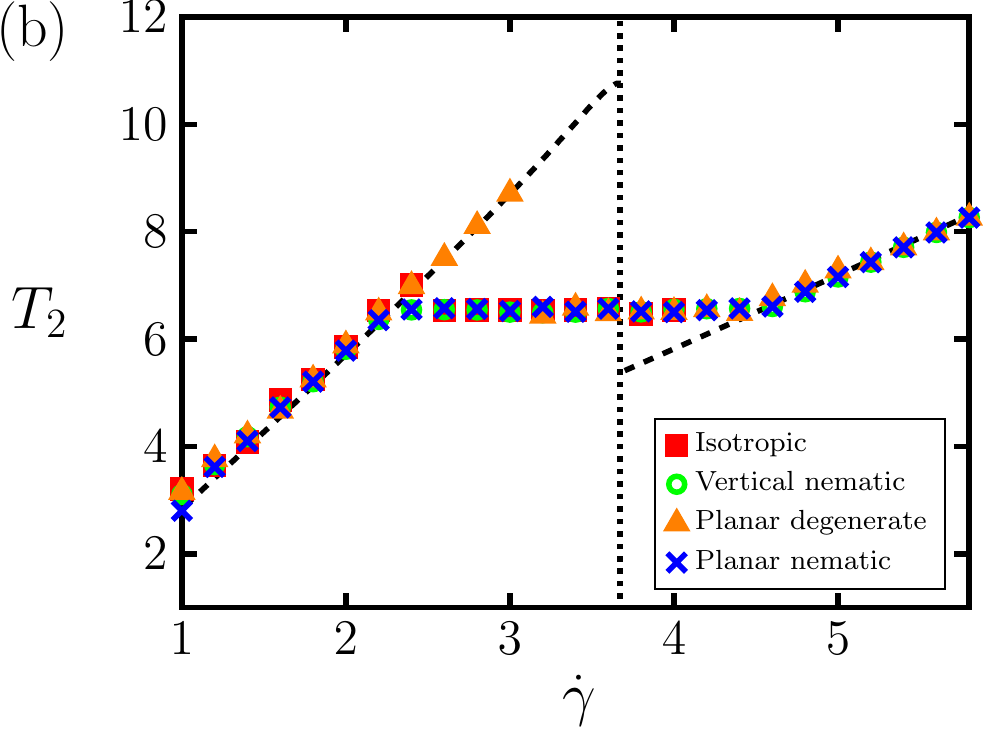}
\caption{(Color online) Influence of the boundary conditions 
on the inhomogeneous flow curves for 
(a) initially isotropic systems at $\Theta=1.20$, $\lambda=0.55$ and (b) initially nematic systems at $\Theta=-0.25$, $\lambda=1.25$. The correlation length is set to $\xi^2 = 10^{-5}$. As a reference the homogeneous flow curves have been included.}
\label{Flow-Boundaries}
\end{figure}

Numerical results for the resulting flow curve of the inhomogeneous systems already discussed
in sect.~\ref{SubSubSec:03.2.1} are presented in fig.~\ref{Flow-Boundaries}. 
We first consider systems sheared from the isotropic state at $\lambda=0.55$, where we found a clear
shear banding instability for fully isotropic boundary conditions [see figs.~\ref{Inhom-Isotropic} and~\ref{Inhom-Flow-Isotropic}]. 
As indicated by the flow curves in fig.~\ref{Flow-Boundaries}(a), similar behavior occurs for other (including nematic) boundary conditions. Indeed, as an analysis of the $\mathbf{Q}$--tensor reveals, all
of the systems form bands (within a range of shear rates $\dot\gamma\approx 0.355$ -- $\dot\gamma=0.385$) with wagging--like oscillations within paranematic regimes at the plates. Outside the
banding region, the systems are characterized by the same value of $T_2$.

Moreover, the value of the "selected" stress within the banding region, $T_2^{sel} \approx 0.59$, is essentially independent of the boundary conditions. The latter only affect the {\em onset} of shear banding upon 
increasing $\dot\gamma$ from low shear rates. Specifically, for the two types of isotropic boundary conditions [eqs.~(\ref{Eq:Bound-Isotropic}) and~(\ref{Eq:Bound-PlanDegenerate})], as well as for nematic ordering within the plane of the plates 
[eq.~(\ref{Eq:Bound-PlanNematic})], the system stays in the homogeneous paranematic state for all $\dot\gamma$ up to the maximum of the flow curve. In contrast, the system with 
vertical alignment at the plates, {\em i.e.}, alignment in the shear gradient ($y$--) direction [see eq.~(\ref{Eq:Bound-VertNematic})], 
forms bands once the value of $T_2^{sel}$ is reached (at $\dot{\gamma} \approx 0.34$). {In this sense, the vertical nematic ordering
favors the occurrence of the W state characterized by oscillations in the flow--gradient plane. For completeness we also note that, upon decreasing $\dot\gamma$ 
from high values, the system goes without hysteresis into the banded state (with the same stress), irrespective of the boundary conditions.

We now turn to the initially nematic case [see fig.~\ref{Flow-Boundaries}(b)]. Upon increasing $\dot\gamma$ from lower values all systems, irrespective of boundary conditions, 
display shear band formation
at shear rates in the range $\dot\gamma\approx 3.0$ -- $\dot\gamma\approx 4.5$; here they break up into a band with kayaking--tumbling dynamics ({\em i.e.}, oscillations
out of the shear plane) surrounded by regions of shear--alignment at the boundaries [see fig.~\ref{Inhom-Nematic} for results with planar nematic boundaries]. 
Further, the stress $T_2$ characterizing the banded state seems to be unique, 
and the boundary conditions only affect the onset of shear banding (upon starting from the low--shear rate branch, where the system is in the KT state).
Specifically, we see from fig.~\ref{Flow-Boundaries}(b) that the onset of banding is "delayed" when we use planar degenerate boundaries [see eq.~(\ref{Eq:Bound-PlanDegenerate})]. These boundary conditions seem to support the KT state, which is understandable as the KT oscillations are out of the shear plane and thus, do involve the plane of the plates. 
Interestingly, the behavior upon decreasing the shear rate from the aligned state is different: in that case, all systems stay in the aligned state until the lower end of the
high--shear rate branch is reached; then they directly jump into the KT state without an intermediate shear banded state.

To summarize, both systems considered in fig.~\ref{Flow-Boundaries} display shear banding irrespective of the detailed nature of the boundary conditions, if the shear rate is increased from low values. The boundary conditions then only influence the "critical" 
shear rate at which the homogeneous state observed at small $\dot\gamma$ breaks up into bands. On the contrary, shear banding upon decreasing $\dot\gamma$ from high values
is only seen in the initially isotropic system. The behavior in the initially nematic system thus depends on the initial conditions, similar to what has
been found in the DJS model~\cite{Adams2008}.
\begin{figure*}
\centering
\includegraphics[height=4.0cm,keepaspectratio]{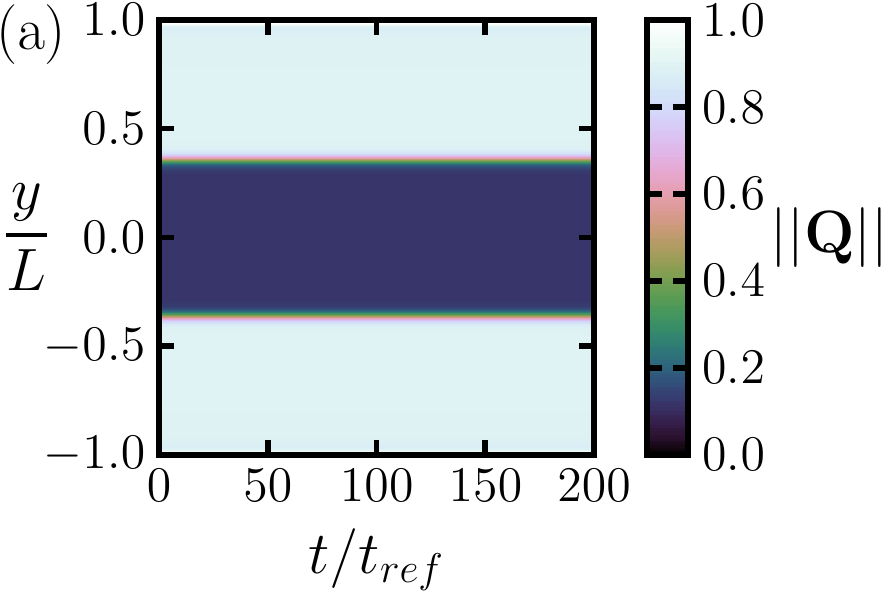}$\quad$
\includegraphics[height=4.0cm,keepaspectratio]{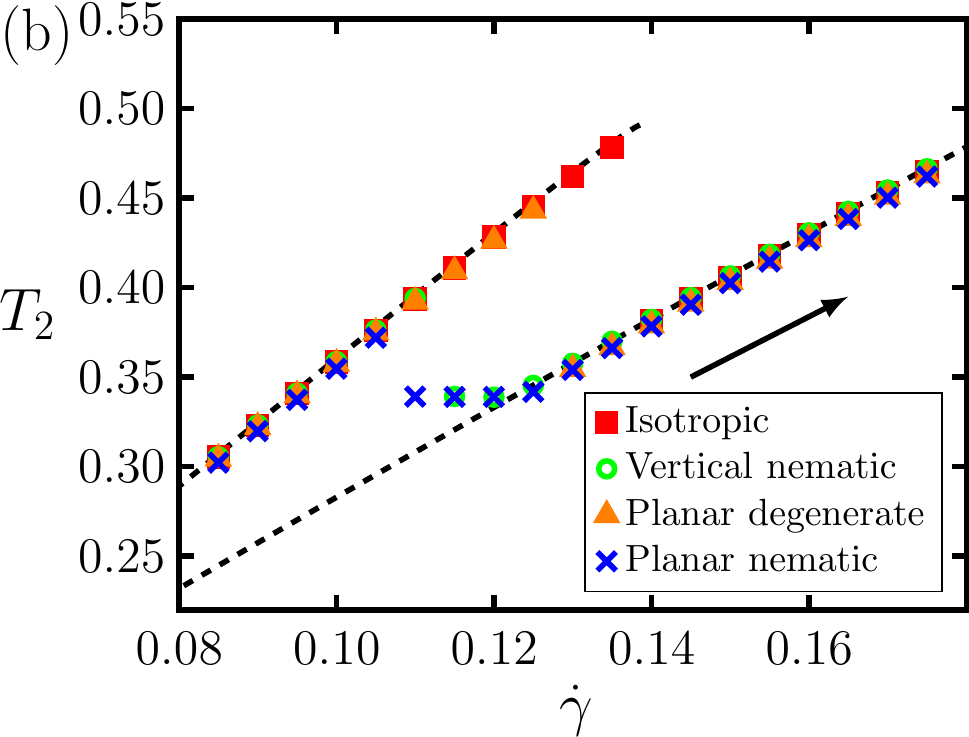}$\quad$
\includegraphics[height=4.0cm,keepaspectratio]{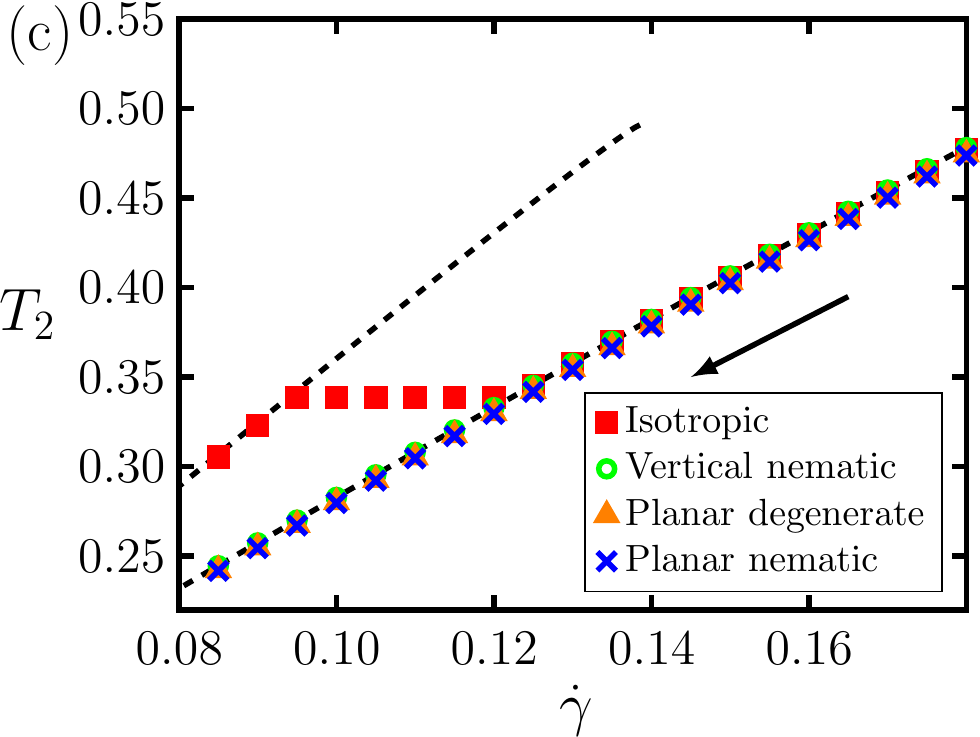}
\caption{(Color online) a) Space--time plot of the norm of the $\mathbf{Q}$--tensor 
for the initially isotropic system at $\Theta=1.20$, $\lambda=1.25$ with vertical nematic boundary conditions. The (average) shear rate and the correlation length
are set to $\dot\gamma=0.115$ and $\xi^2=10^{-5}$, respectively. b) Inhomogeneous flow curves for the system with different boundary conditions upon starting from
low shear rates. c) Same as b), but starting from high shear rates}.
\label{Iso-Boundaries}
\end{figure*}

Finally, we come back to a system which we already considered briefly in sect.~\ref{SubSubSec:03.1.1}, that is, the case $\Theta=1.20$ and $\lambda=1.25$.
This initially isotropic system is characterized by a discontinuous homogeneous flow curve [related to the transition from paranematic to shear--aligned state, see
fig.~\ref{Hom-Isotropic}(c)], but does not form clear shear bands when using disordered boundary conditions [see eqs.~(\ref{Eq:Bound-Isotropic}) or (\ref{Eq:Bound-PlanDegenerate})]. 
Interestingly, this changes when we use "nematic" boundary conditions with alignment either in the plane of the plates  [eq.~(\ref{Eq:Bound-PlanNematic})] or along the vorticity direction [eq.~(\ref{Eq:Bound-VertNematic})].
As an illustration, we present in fig.~\ref{Iso-Boundaries}(a) the spatiotemporal behavior of the $\mathbf{Q}$--tensor at $\xi^2=10^{-5}$, with planar boundaries and an imposed shear rate $\dot{\gamma}=0.115$. The plot reveals a band with paranematic ordering within the outer regions where the system is shear--aligned. Figure~\ref{Iso-Boundaries}(b) shows results for corresponding flow curves. Consistent with the previous observations, we observe a plateau
in $T_2(\dot\gamma)$, related to shear band formation, only with nematic boundary conditions, whereas the disordered boundary conditions yield an abrupt change between two homogeneous 
states. Yet a different behavior is found when we decrease the shear rate from high values, see fig.~\ref{Iso-Boundaries}(c): In this case, stable shear bands are found only for fully isotropic boundary conditions characterized by three--dimensional disorder.
\section{Concluding remarks}
\label{Sec:04}
In this paper we have investigated the occurrence of shear bands in nematogenic fluids based on the mesoscopic Doi--Hess theory for the orientational order parameter tensor 
$\mathbf{Q}$ coupled to the shear stress $\mathbf{T}$. We have focused on instabilities along the gradient direction (gradient banding).
Studying sheared systems with different (isotropic or nematic) equilibrium states, different tumbling parameters ({\em i.e.}, aspect ratios), and different orientational boundary conditions, our results
reveal complex shear banding behavior whose characteristics can be significantly different from that seen in other models, such as the DJS model (where the dynamical variable is the shear stress alone).
In most cases considered, the shear bands involve
oscillatory (but not chaotic) orientational motion in certain regions of space. In terms of parameters, our study  extents earlier investigations focusing on the 
isotropic--nematic transition~\cite{Olmsted1999} or the rheochaotic regime~\cite{Chakrabarti2004,Das2005}.}

"Classical" behavior characterized by the S--shaped flow curve (obtained from the homogeneous solutions) and a unique value of the stress in the banded regime (such as in the DJS model) is found only in one case, namely an initially isotropic
system with relatively small tumbling parameter. This system displays bands with wagging--like motion in the inner part and steady alignment close to the boundaries. In all other cases,
the homogeneous flow curves display  a discontinuity (rather than the S--shape). The observed band formation then depends on the orientational boundary conditions 
in the sense that certain boundary conditions can support or hinder the formation of shear bands. Moreover, in one case we found a strong dependence
on the pathway of the shear protocol. These observations suggest that the mechanisms of shear banding and stress selection are more complex than in 
simpler (such as the DJS) models~\cite{Olmsted2000,Lu2000}.

An important question arising from the present work concerns the relation between the parameter sets ($\theta$, $\lambda$, $\dot\gamma$) considered here and the system's "phase" 
diagram under shear. In many nematogenic systems the relevant variable for the isotropic--nematic transition is the 
concentration, which does not occur explicitly in our approach. However, there is an implicit concentration dependence through the free energy functional (specifically, the prefactor $\theta$ of the quadratic term).
Based on that dependence, we have proposed in an earlier study~\cite{Strehober2013} a phase diagram in the concentration--shear rate plane, which qualitatively resembles
earlier results~\cite{Dhont2008,Olmsted1999}. In particular, the diagram reproduces the shear--induced shift of the isotropic--nematic transition towards smaller concentrations, as well as a critical shear rate beyond which
the transition becomes continuous.
Based on ref.~\cite{Strehober2013} we find
that the initially isotropic state with small tumbling parameter (with $\theta=1.20$, $\lambda=0.55$) considered here is very close to the critical point
(see fig.~7 in~\cite{Strehober2013}). The appearance of shear banding (in gradient direction) 
in this situation is indeed consistent with earlier predictions for systems of colloidal rods, such as suspensions of $fd$--viruses~\cite{Dhont2008,Dhont2002}. The other parameter sets considered in the present work lie deep in the nematic phase of the concentration--shear rate phase diagram.

A further interesting point concerns the occurrence of shear bands in vorticity direction, a scenario which we have implicitly ruled out by focusing on inhomogeneities along the $y$--direction alone.
Indeed, vorticity banding has been predicted to occur in colloidal rod ($fd$--virus) suspensions at conditions {\em within} the isotropic--nematic spinodal under shear~\cite{Dhont2008}.
Conceptually, vorticity--banded states are characterized by the same shear rate (rather than same shear stress as gradient--banded states)~\cite{Olmsted2008}. With this background,
the shape of the (homogeneous) flow curves obtained here (namely those characterized by discontinuities and large hysteresis) may be taken as an indication
for vorticity banding. This point certainly warrants further investigation. Another important direction is the investigation of (binary) mixture system
which have even more complex dynamics already in the homogeneous case~\cite{LugoFrias2016}. Work in these directions is in progress.

\section*{Acknowledgments}
We acknowledge financial support from the Deutsche For\-schungsgemeinschaft through the Research Training Group 1558, project B3. The authors would also like to thank S. Heidenreich for very fruitful discussions.

\section*{Appendix: Dimensionless form of the dynamical equations}
In order to rewrite eqs.~(\ref{Eq:Hess-Qtensor}) and~(\ref{Eq:Hess-Stensor}) in a dimensionless form, we first introduce (see ref.~\cite{LugoFrias2016})
the scaled order--parameter tensor $\tilde{\mathbf{Q}}$ and the scaled free energy $\tilde{\mathscr{F}^{or}}$ via the relations 
\begin{align}
\label{Eq:Scaling}
\tilde{\mathbf{Q}} = \frac{\mathbf{Q}}{\mu_3^k} \quad \text{and} \quad \tilde{\mathscr{F}^{or}_{h}} = \frac{\mathscr{F}^{or}_{h}}{\mathscr{F}^{or}_{ref}} \,.
\end{align}
Here, $\mu_3^k$ is the value of the uniaxial order parameter at the isotropic--nematic phase transition, $\mu_3^k= {\sqrt{6} B}/{12 C}$, whereas 
$\mathscr{F}^{or}_{ref} = 2 C {\mu_3^k}^4$ corresponds to a reference value of the free energy in equilibrium (for more details, see~\cite{LugoFrias2016}). 
Using eqs.~(\ref{Eq:Scaling}) the scaled orientational free energy becomes
\begin{align}
\label{Eq:Scaled-Energy}
\tilde{\mathscr{F}^{or}_{h}} = \frac{\Theta}{2}\!\left(\tilde{\mathbf{Q}}\!:\!\tilde{\mathbf{Q}} \right) - \sqrt{6}\! \left( \tilde{\mathbf{Q}} \!\cdot\!\tilde{\mathbf{Q}} \right)\!:\!\tilde{\mathbf{Q}} + \frac{1}{2}\! \left(\tilde{\mathbf{Q}} \!:\!\tilde{\mathbf{Q}}\right)^2 \,,
\end{align}
where $\Theta=24\,AC/B^2$. Microscopically, 
$\Theta$ depends on the number density and the molecular aspect ratio~\cite{LugoFrias2016}. Specifically, 
for a given aspect ratio, $\Theta$ changes sign (from positive to negative) as the concentration of the system increases and the isotropic--nematic phase transition takes place.
The final expression of the scaled orientational free energy in eq.~(\ref{Eq:Scaled-Energy}) corresponds exactly to the one in earlier studies (see e.g.~\cite{Rienacker2002,Heidenreich2009}). However, because of the definition of $\mu_3^k$ the scaling of the subsequent variables is modified by a constant factor.

To proceed, we introduce (following refs.~\cite{LugoFrias2016,Heidenreich2009}) the scaled spatial coordinate
$\tilde{\mathbf{r}} = {\mathbf{r}}/{L}$ (where $L$ is half of the separation between the plates, see fig.~\ref{Fig:Sketch})
and the scaled time $\tilde{t} = {t}/{t_{ref}}$ with
$t_{ref} = \tau_q {\mu_3^k}^2/ \mathscr{F}^{or}_{ref}$. Using these dimensionless variables together with eq.~(\ref{Eq:Scaled-Energy}) in eqs.~(\ref{Eq:Hess-Qtensor}) and~(\ref{Eq:Hess-Stensor}) we obtain
eqs.~(\ref{Eq:Exp-Qtensor}) and~(\ref{Eq:Exp-Stensor}) in the main text. 
The scaled correlation length and velocity field are $\tilde{\xi}^2 = {\mu_3^k}^2 \xi^2/(L^2 \mathscr{F}^{or}_{ref})$
and $\tilde{\mathbf{v}} = \mathbf{v}/L\dot{\gamma}$, respectively. Further, the nonlinear source term [see eq.~(\ref{Eq:Exp-Qtensor})] becomes
\begin{align}
\label{Eq:Scaled-Source}
\tilde{\mathbf{H}}(\tilde{\mathbf{Q}},\tilde{\mathbf{v}}) =2 \tilde{\dot{\gamma}}\!\stl{\tilde{\mathbf{\Omega}}\!\cdot\!\tilde{\mathbf{Q}}}\! + 2 \sigma \tilde{\dot{\gamma}}\!\stl{\tilde{\mathbf{\Gamma}}\!\cdot\!\tilde{\mathbf{Q}}}\! + \sqrt{2}\tilde{\lambda} \tilde{\mathbf{\Gamma}} - \tilde{\mathbf{\Phi}}'\,,
\end{align}
where $\tilde{\dot{\gamma}}=\dot{\gamma}t_{ref}$, $\tilde{\lambda}=\lambda/\mu_3^k$, and 
the derivative of the orientational free energy is 
\begin{align}
\label{Eq:Scaled-DevEnergy}
\tilde{\mathbf{\Phi}}' = \Theta \mathbf{\tilde{Q}} - 3\sqrt{6} \stl{\mathbf{\tilde{Q}} \cdot \mathbf{\tilde{Q}}}  + 2 (\mathbf{\tilde{Q}}:\mathbf{\tilde{Q}}) \cdot \mathbf{\tilde{Q}}\,.
\end{align}
For the present flow
geometry [see fig.~\ref{Fig:Sketch}], the vorticity and deformation tensors take the form $\mathbf{\tilde{\Omega}} = (1/2)(\hat{\mathbf{e}}^x \hat{\mathbf{e}}^y - \hat{\mathbf{e}}^y \hat{\mathbf{e}}^x)$ and~$\mathbf{\tilde{\Gamma}} = (1/2)(\hat{\mathbf{e}}^x \hat{\mathbf{e}}^y + \hat{\mathbf{e}}^y \hat{\mathbf{e}}^x)$, respectively. 

Finally, the stress tensor~(\ref{Eq:Stress-Tensor}) is expressed in terms of the scaled variables as
\begin{equation}
\label{Eq:Scaled-Stress}
\tilde{\mathbf{T}} = \frac{\mathbf{T}}{p_{kin} \mathscr{F}^{or}_{ref}} = - \tilde{p}\,\mathbb{I} + 2\tilde{\eta}_{iso} \tilde{\dot{\gamma}} \tilde{\mathbf{\Gamma}} + \stl{\tilde{\mathbf{T}}_{al}}\,,
\end{equation}
where $p_{kin} =\rho k_B T/m$ is the prefactor of the alignment contribution,
$\stl{{\mathbf{T}}_{al}}$, in eq.~(\ref{Eq:Stress-Alignment}) and $\tilde{\eta}_{iso} = \eta_{iso}/(p_{kin} t_{ref}\mathscr{F}^{or}_{ref})$. The alignment contribution to the stress then becomes
\begin{align}
\label{Eq:Scaled-Stress-Alignment}
\stl{\tilde{\mathbf{T}}_{al}} = &\, \sqrt{2} \tilde{\lambda} \tilde{\mathbf{\Phi}}' - \sqrt{2} \tilde{\lambda} \tilde{\xi}^2 \tilde{\nabla}^2 \tilde{\mathbf{Q}} \nonumber \\ &\, - 2\sigma  \stl{\tilde{\mathbf{Q}} \cdot \tilde{\mathbf{\Phi}}'} + 2\sigma \tilde{\xi}^2 \stl{\tilde{\mathbf{Q}} \cdot \tilde{\nabla}^2 \tilde{\mathbf{Q}}} \,.
\end{align}

To illustrate the connection between our scaling procedure and real systems, consider solutions of {\em fd}--virus with a Maier-Saupe order parameter at coexistence $S_k \sim 0.5$~\cite{Purdy2003} and 
a cholesteric pitch $\sim 10^{-6} m$~\cite{Dogic2000} in a Couette cell of $L = 10^{-3} m$~\cite{Fardin2012}. Assuming a typical relaxation time $\tau_q \sim 0.01s$ and using the relation
$\mu_3^k = \sqrt{5} S_k$, the values of the scaled variables are $\tilde{\lambda} \approx 0.48 - 0.69$,  $\tilde{\dot{\gamma}} \approx 0.7-2.0$ and $\tilde{\xi}^2 \approx 1.25\times 10^{-5}$.

%
\bibliographystyle{epj}
\bibliography{LF_Reinken_Klapp_verFinal.bib}

\begin{thebibliography}{54}

\bibitem{Fielding2007}
S.M. Fielding, Soft Matter \textbf{3}, 1262 (2007)

\bibitem{Dhont2008}
J.K. Dhont, W.J. Briels, Rheol. Acta \textbf{47}, 257 (2008)

\bibitem{LopezGonzalez2004}
M.~Lopez-Gonzalez, W.~Holmes, P.~Callaghan, P.~Photinos, Phys. Rev. Lett.
  \textbf{93}, 268302 (2004)

\bibitem{Decruppe1995}
J.~Decruppe, R.~Cressely, R.~Makhloufi, E.~Cappelaere, Colloid Polym. Sci.
  \textbf{273}, 346 (1995)

\bibitem{Chen1992}
L.~Chen, C.~Zukoski, B.~Ackerson, H.~Hanley, G.~Straty, J.~Barker, C.~Glinka,
  Phys. Rev. Lett. \textbf{69}, 688 (1992)

\bibitem{Chikkadi2014}
V.~Chikkadi, D.~Miedema, M.~Dang, B.~Nienhuis, P.~Schall, Phys. Rev. Lett.
  \textbf{113}, 208301 (2014)

\bibitem{Goveas2001}
J.~Goveas, P.~Olmsted, Eur. Phys. J. E \textbf{6}, 79 (2001)

\bibitem{Spenley1996}
N.~Spenley, X.~Yuan, M.~Cates, J. Phys. II \textbf{6}, 551 (1996)

\bibitem{Olmsted2000}
P.~Olmsted, O.~Radulescu, C.Y. Lu, J. Rheol. \textbf{44}, 257 (2000)

\bibitem{Tao2005}
Y.G. Tao, W.~den Otter, W.~Briels, Phys. Rev. Lett. \textbf{95}, 237802 (2005)

\bibitem{Ripoll2008}
M.~Ripoll, P.~Holmqvist, R.G. Winkler, G.~Gompper, J.K.G. Dhont, M.P. Lettinga,
  Phys. Rev. Lett. \textbf{101}, 168302 (2008)

\bibitem{Olmsted1999}
P.D. Olmsted, C.Y.D. Lu, Phys. Rev. E \textbf{60}, 4397 (1999)

\bibitem{Hess1994}
O.~Hess, S.~Hess, Phys. A \textbf{207}, 517 (1994)

\bibitem{Rienacker2000}
G.~Rien\"acker, \emph{Orientational dynamics of nematic liquid crystals in a
  shear flow} (Shaker Verlag, Aachen, 2000)

\bibitem{Rienacker2002}
G.~Rien\"acker, M.~Kr\"oger, S.~Hess, Phys. Rev. E \textbf{66}, 040702 (2002)

\bibitem{Strehober2013}
D.A. Strehober, H.~Engel, S.H.L. Klapp, Phys. Rev. E \textbf{88}, 012505 (2013)

\bibitem{Olmsted1991}
P.D. Olmsted, Ph.D. thesis, University of Illinois, Urbana-Champaign (1991)

\bibitem{Roux1995}
D.C. Roux, J.F. Berret, G.~Porte, E.~Peuvrel-Disdier, P.~Lindner,
  Macromolecules \textbf{28}, 1681 (1995)

\bibitem{Lettinga2005}
M.~Lettinga, Z.~Dogic, H.~Wang, J.~Vermant, Langmuir \textbf{21}, 8048 (2005)

\bibitem{Chakraborty2010}
D.~Chakraborty, C.~Dasgupta, A.K. Sood, Phys. Rev. E \textbf{82}, 065301 (2010)

\bibitem{Chakrabarti2004}
B.~Chakrabarti, M.~Das, C.~Dasgupta, S.~Ramaswamy, A.~Sood, Phys. Rev. Lett.
  \textbf{92}, 055501 (2004)

\bibitem{Das2005}
M.~Das, B.~Chakrabarti, C.~Dasgupta, S.~Ramaswamy, A.K. Sood, Phys. Rev. E
  \textbf{71}, 021707 (2005)

\bibitem{Dhont2002}
J.K.G. Dhont, M.P. Lettinga, Z.~Dogic, T.A.J. Lenstra, H.~Wang, S.~Rathgeber,
  P.~Carletto, L.~Willner, H.~Frielinghaus, P.~Lindner, Farad. Discuss.
  \textbf{123}, 157 (2003)

\bibitem{deGennes1993}
P.G. de~Gennes, \emph{The physics of liquid crystals} (Clarendon Press, Oxford,
  1993)

\bibitem{Hess2015}
S.~Hess, \emph{Tensors for Physics} (Springer Intl., Switzerland, 2015)

\bibitem{LugoFrias2016}
R.~Lugo-Frias, S.H.L. Klapp, J. Phys. Condens. Matter  (2016)

\bibitem{Oseen1933}
C.~Oseen, Trans. Faraday Soc. \textbf{29}, 883 (1933)

\bibitem{Frank1958}
F.C. Frank, Discuss. Faraday Soc. \textbf{25}, 19 (1958)

\bibitem{Singh1985}
Y.~Singh, K.~Singh, Phys. Rev. A \textbf{33}, 3481 (1985)

\bibitem{Singh1986}
K.~Singh, Y.~Singh, Phys. Rev. A \textbf{34}, 548 (1986)

\bibitem{Singh1987}
K.~Singh, Y.~Singh, Phys. Rev. A \textbf{35}, 3535 (1987)

\bibitem{Taylor1923}
G.~Taylor, Proceedings of the Royal Society of London. Series A, Containing
  Papers of a Mathematical and Physical Character \textbf{103}, 58 (1923)

\bibitem{Hess1975}
S.~Hess, Z. Naturforsch. \textbf{30a}, 728 (1975)

\bibitem{Pardowitz1980}
I.~Pardowitz, S.~Hess, Phys. A \textbf{100}, 540 (1980)

\bibitem{Heidenreich2009}
S.~Heidenreich, Ph.D. thesis, TU Berlin (2009)

\bibitem{Borgmeyer1995}
C.P. Borgmeyer, S.~Hess, J. Non-Equilib. Thermodyn. \textbf{20}, 359 (1995)

\bibitem{Hess1976}
S.~Hess, Z. Naturforsch. \textbf{31a}, 1034 (1976)

\bibitem{deGroot1983}
S.R. De~Groot, P.~Mazur, \emph{Non-equilibrium thermodynamics} (Dover, 1983)

\bibitem{Ganapathy2008}
R.~Ganapathy, S.~Majumdar, A.K. Sood, Phys. Rev. E \textbf{78}, 021504 (2008)

\bibitem{Hess2004}
S.~Hess, M.~Kr\"oger, J. Phys. Condens. Matter \textbf{16}, S3835 (2004)

\bibitem{Hess2005}
S.~Hess, M.~Kr\"oger, \emph{in Computer Simulations of Liquid Crystals and
  Polymers} (Springer Verlag, 2005), pp. 295--333

\bibitem{Kaiser1992}
P.~Kaiser, W.~Wiese, S.~Hess, J. Non-Equilib. Thermodyn. \textbf{17}, 153
  (1992)

\bibitem{Press1996}
W.H. Press, S.A. Teukolsky, W.T. Vetterling, B.P. Flannery, \emph{Numerical
  recipes in C}, Vol.~2 (Cambridge university press, Cambridge, 1996)

\bibitem{Sheng1982}
P.~Sheng, Phys. Rev. A \textbf{26}, 1610 (1982)

\bibitem{Jerome1991}
B.~Jerome, Rep. Prog. Phys. \textbf{54}, 391 (1991)

\bibitem{Ruths1996}
M.~Ruths, S.~Steinberg, J.N. Israelachvili, Langmuir \textbf{12}, 6637 (1996)

\bibitem{Manyuhina2010}
O.~Manyuhina, A.M. Cazabat, M.B. Amar, Europhys. Lett. \textbf{92}, 16005
  (2010)

\bibitem{Radulescu2000}
O.~Radulescu, P.~Olmsted, J. Non-Newton. Fluid \textbf{91}, 143 (2000)

\bibitem{Adams2008}
J.~Adams, S.~Fielding, P.~Olmsted, Journal of Non-Newtonian Fluid Mechanics
  \textbf{151}, 101 (2008)

\bibitem{Lu2000}
C.Y.D. Lu, P.D. Olmsted, R.~Ball, Phys. Rev. Lett. \textbf{84}, 642 (2000)

\bibitem{Olmsted2008}
P.D. Olmsted, Rheol. Acta \textbf{47}, 283 (2008)

\bibitem{Purdy2003}
K.R. Purdy, Z.~Dogic, S.~Fraden, A.~R\"uhm, L.~Lurio, S.G.J. Mochrie, Phys.
  Rev. E \textbf{67}, 031708 (2003)

\bibitem{Dogic2000}
Z.~Dogic, S.~Fraden, Langmuir \textbf{16}, 7820 (2000)

\bibitem{Fardin2012}
M.~Fardin, T.~Ober, V.~Grenard, T.~Divoux, S.~Manneville, G.~McKinley,
  S.~Lerouge, Soft Matter \textbf{8}, 10072 (2012)

\end{thebibliography}

\end{document}